\documentclass[fleqn,usenatbib]{mnras}

\usepackage{newtxtext,newtxmath}

\usepackage[T1]{fontenc}

\DeclareRobustCommand{\VAN}[3]{#2}
\let\VANthebibliography\thebibliography
\def\thebibliography{\DeclareRobustCommand{\VAN}[3]{##3}\VANthebibliography}

\usepackage{graphicx}	
\usepackage{amsmath}	

\newcommand{\msolt}{M$_\odot$}
\newcommand{\pcsqt}{pc$^{-2}$}

\newcommand{\be}{\begin{equation}}
\newcommand{\ee}{\end{equation}}
\newcommand{\ie}{\emph{i.e.}, }
\newcommand{\eg}{\emph{e.g.}, }

\title[Spiral Arms are Metal Freeways]{Spiral Arms are Metal Freeways: Azimuthal Gas-Phase Metallicity Variations in Simulated Cosmological Zoom-in Flocculent Disks}

\author[M. E. Orr et al.]{
Matthew E. Orr,$^{1,2}$\thanks{E-mail: matt.orr@rutgers.edu}
Blakesley Burkhart,$^{1,2}$
Andrew Wetzel,$^3$
Philip F. Hopkins,$^{4}$
Ivanna A. Escala,$^{5,6}$\thanks{Carnegie-Princeton Fellow}\newauthor
\textcolor{black}{Allison L. Strom,$^{5,6,7}$\textdagger}
Paul F. Goldsmith,$^{8}$
Jorge L. Pineda,$^{8}$ 
Christopher C. Hayward,$^2$
\textcolor{black}{Sarah R. Loebman$^{9}$}
\\
$^{1}$Department of Physics and Astronomy, Rutgers University, 136 Frelinghuysen Road, Piscataway, NJ 08854, USA\\
$^{2}$Center for Computational Astrophysics, Flatiron Institute, 162 Fifth Avenue, New York, NY 10010, USA\\
$^3$Department of Physics \& Astronomy, University of California, Davis, CA 95616, USA\\
$^{4}$TAPIR, Mailcode 350-17, California Institute of Technology, Pasadena, CA 91125, USA\\
$^5$The Observatories of the Carnegie Institution for Science, 813 Santa Barbara Street, Pasadena, CA 91101, USA\\
$^6$Department of Astrophysical Sciences, Princeton University, 4 Ivy Lane, Princeton, NJ 08544, USA\\
$^{7}$Department of Physics and Astronomy and CIERA, Northwestern University, 2145 Sheridan Road, Evanston, IL 60208, USA\\
$^8$Jet Propulsion Laboratory, California Institute of Technology, 4800 Oak Grove Drive, Pasadena, CA 91109-8099, USA\\
$^9$Department of Physics, University of California, Merced, CA 95343, USA\\
}

\date{Accepted XXX. Received YYY; in original form ZZZ}

\pubyear{2022}

\begin{document}
\label{firstpage}
\pagerange{\pageref{firstpage}--\pageref{lastpage}}
\maketitle

\begin{abstract}
We examine the azimuthal variations in gas-phase metallicity profiles in simulated Milky Way mass disk galaxies from the Feedback in Realistic Environments (FIRE-2) cosmological zoom-in simulation suite, which includes a sub-grid turbulent metal mixing model. We produce spatially resolved maps of the disks at $z \approx 0$ with pixel sizes ranging from 250 to 750~pc, analogous to modern integral field unit (IFU) galaxy surveys, mapping the gas-phase metallicities in both the cold \& dense gas and the ionized gas correlated with HII regions.  
We report that the spiral arms alternate in a pattern of metal rich and metal poor relative to the median metallicity on the order of $\lesssim 0.1$~dex, appearing generally in this sample of flocculent spirals.
The pattern persists even in a simulation with different strengths of metal mixing, indicating that the pattern emerges from physics above the sub-grid scale.
Local enrichment does not appear to be the dominant source of the azimuthal metallicity variations at $z \approx 0$: there is no correlation with local star formation on these spatial scales.
Rather, the arms are moving inwards and outwards relative to each other, carrying their local metallicity gradients with them radially before mixing into the larger-scale interstellar medium.  
We propose that the arms act as freeways channeling relatively metal poor gas radially inwards, and relatively enriched gas radially outwards.
\end{abstract}



\section{Introduction}

Spiral arms are a common feature of disk galaxies at late times, such as those in our own Milky Way.  Though their origins and physical causes remain varied and uncertain \citep{Goldreich1965a, Schwartz1984, Dobbs2010, Purcell2011, DOnghia2013}, and their long-term stability is not a given \citep{Sellwood2011}, it is clear that in these galaxies circumgalactic gas generally joins up with the disks in their outskirts and migrates inwards \citep{Trapp2022}, rather than impacting and accreting into the galaxies at a variety of radii \citep{Peroux2020, Stern2021,Hafen2022}.

Another typical feature of relatively isolated spirals at late times, which are not undergoing a major merger, is that their disks are in rough dynamical equilibrium.  
Their star formation rates are smooth, the gas disks are in vertical hydrostatic equilibrium, and they are marginally stable ($Q\approx 1$) against gravitational fragmentation and collapse \citep{Leroy2008, Kim2015, Krumholz2016, Gurvich2020, Orr2020, Wang2020, Ostriker2022}.  
In galaxies that are able to maintain smooth star formation rates (SFRs) for at least a few gas depletion times (defined as $\Sigma_{\rm gas}/\Sigma_{\rm SFR} \equiv t_{\rm dep}$, see \eg \citealt{Kennicutt1998, Leroy2008, Burkhart2019}), gas transport must proceed apace to replace gas as it is slowly astrated.  
This requires a net inspiraling velocity of $v_{\rm in} \sim R_{\rm disk}/t_{\rm dep} \sim 10\; {\rm kpc}/(1-3\; {\rm Gyr}) \sim 3-10$ km/s in the gas disk.  
Such transport velocities are one to two orders of magnitude lower than typical orbital velocities in MW-mass galaxies ($\sim$100-300 km/s, \citealt{DeBlok2008, Lelli2016}), and roughly comparable to the turbulent velocity dispersions in the dense molecular gas \citep{Sun2020} and the thermal velocity dispersions in either the warm neutral medium ($T\sim 6000-8000$~K, $c_s \approx 6.3-7.3$~km/s) or the warm ionized medium ($T\sim10^4$~K, $c_s \approx 12$~km/s).  That the velocity of the transport is so much smaller than the orbital velocities and commensurate with the (variously sourced) dispersions makes it very difficult to observationally disentangle. 
Significant observational effects such as beam smearing are at play (see \citealt{Zhou2017} for the importance of removing beam smearing while quantifying spatially resolved velocity dispersions).
Confounding gas kinematics and dynamics from, \eg disk warps or holes make subtracting these other features intrinsically difficult and isolating the gas transport directly from the velocity structure in line emission from the gas challenging. Cases in point: \citet{Schmidt2016} found that the radial inflow of HI in disk outskirts did, in fact, roughly match measured SFRs. More recent work by \citet{DiTeodoro2021} used publicly available HI data to quantify radial atomic gas motion in nearby spirals with modeling that included treatment of disk warps. They found the magnitude of radial motions to be a few km/s though it was unclear if the motions amounted to systematic radial inward or outward motions.

However, the velocity structure in line emission from the molecular, atomic and ionized interstellar medium (ISM) phases is not the only potential window for understanding gas transport in spiral galaxies.  Indeed, radial gas transport in disks has been invoked also in interpreting turbulent velocity dispersions.  Relating again to maintaining the gas disks against gravitational instability and collapse, the conversion of orbital gravitational potential energy to turbulence through gravitoturbulent instabilities as gas spirals inwards has been seen as a non-stellar-feedback source of support \citep{Krumholz2018, Forbes2022, Ginzburg2022}.  
Assuming that even only a few percent of the gravitational potential energy is converted, it is expected that the turbulence driven by mass transport $\sigma_{\rm grav} \approx \psi v_c v_r \approx 0.01 ( \sim$$100-300\; {\rm km/s})(\sim$$3-10\; {\rm km/s}) \approx 3-30 \;{\rm km/s}$ \citep[see \S~4.2 of ][for a short derivation of this scaling and comparison with feedback driven turbulence]{Orr2020}, alone perhaps enough to produce the velocity dispersions observed. In reality we know that mass transport \emph{and} feedback are occurring in star-forming spirals, complicating our ability to use turbulent velocity dispersions to strongly constrain radial mass transport \citep{Yu2021}.

On the other hand, chemical abundance profiles, \ie metallicity gradients, have been held up as integral parts in understanding the assembly histories and the properties of \emph{in situ} star formation in galaxies \citep{Lacey1985, Chiappini2001, Collacchioni2020, Sharda2021}.  
Recent IFU survey work with MaNGA by \citet{Belfiore2017} has shown that as star-forming galaxies build up their disks, relatively strong ($\sim$0.025 dex/kpc) radial gradients in gas-phase metallicity arise.  From the perspective of modern simulations, \citet{Bellardini2021} followed the evolution of gas-phase metallicity gradients (and average scatter) in the FIRE-2 simulations, highlighting the steepening of the gas-phase gradients following the era of disk formation ($z \approx 1.5$). Similarly, \citet{Ma2017a} identified disk formation in FIRE-1 as the point at which negative gas-phase metallicity gradients arose.  The causes of metallicity gradients are varied, but generally observational evidence and theoretical models point towards a balance of inside-out growth (\ie stable, radial SFR profiles \citealt{Ellison2018}) and metal-poor gas accretion in the galactic outskirts (which then migrates inwards). Indeed, work by \citet{Bellardini2022} found that the strength of stellar radial metallicity gradients in FIRE-2 spirals were highly correlated with stellar radial velocity dispersions, suggesting the importance of radial mixing. Additionally on the theoretical side, \citet{Sharda2021} has developed a first-principles model showing the relative importance of metal production, loss, and, critically, gas transport in the development of metallicity gradients in galaxies.  

Work with MaNGA by \citet{Kreckel2019} has investigated azimuthal metallicity variations of HII regions in nearby spirals, finding a weak dependence on local H$\alpha$ luminosity (\ie presumably star formation) and, intriguingly, some correspondence with spiral structures themselves. This motivates the idea that coherent azimuthal metallicity variations along spiral arms might be tied to local enrichment or bulk gas flows, and thus provides a unique perspective in understanding ISM kinematics and evolution.

Given the difficulty in observationally quantifying both in-plane bulk gas motions and spatially resolved metal abundances, we here turn to isolated and cosmological zoom-in galaxy simulations.  
The ability of modern high-resolution galaxy simulations to resolve a multiphase ISM, follow the internal gas dynamics, and include a detailed treatment of star formation with feedback/enrichment physics  \citep{Wetzel2016, Hopkins2018:fire, Agertz2021, Dubois2021} can allow for in-depth studies of the connections between (galactic) disk accretion, star formation, stellar feedback, chemical enrichment, and mass transport \citep{Ma2017, Escala2018, Orr2020, Bellardini2021, Trapp2022}. 
In this paper, we will explore the azimuthal gas-phase metallicity variations in six FIRE-2 Milky Way-mass disk galaxies \citep{Wetzel2016,Hopkins2018:fire} through a lens focused on observables to understand the connections between gas kinematics and dynamics, local enrichment/stellar feedback, and gas-phase metallicity.  
Primarily, we produce maps of the galaxies on 250~pc scales mimicking IFU surveys capable of measuring spatially resolved gas-phase abundances and kinematics in galaxies in nebular HII and cold dense gas like MaNGA or PHANGs \citep{Bundy2015, Leroy2021}.  
We will explore the effects of the turbulent metal diffusion model, as implemented by FIRE-2, on the disk scale metallicity variations.  
We will also interpret the ability of a simple model for gas transport to explain the magnitude of the metallicity variations observed in the FIRE-2 spirals sample.

\section{Simulations \& Methods}\label{sec:methods}
\begin{figure}
	\includegraphics[width=0.48\textwidth]{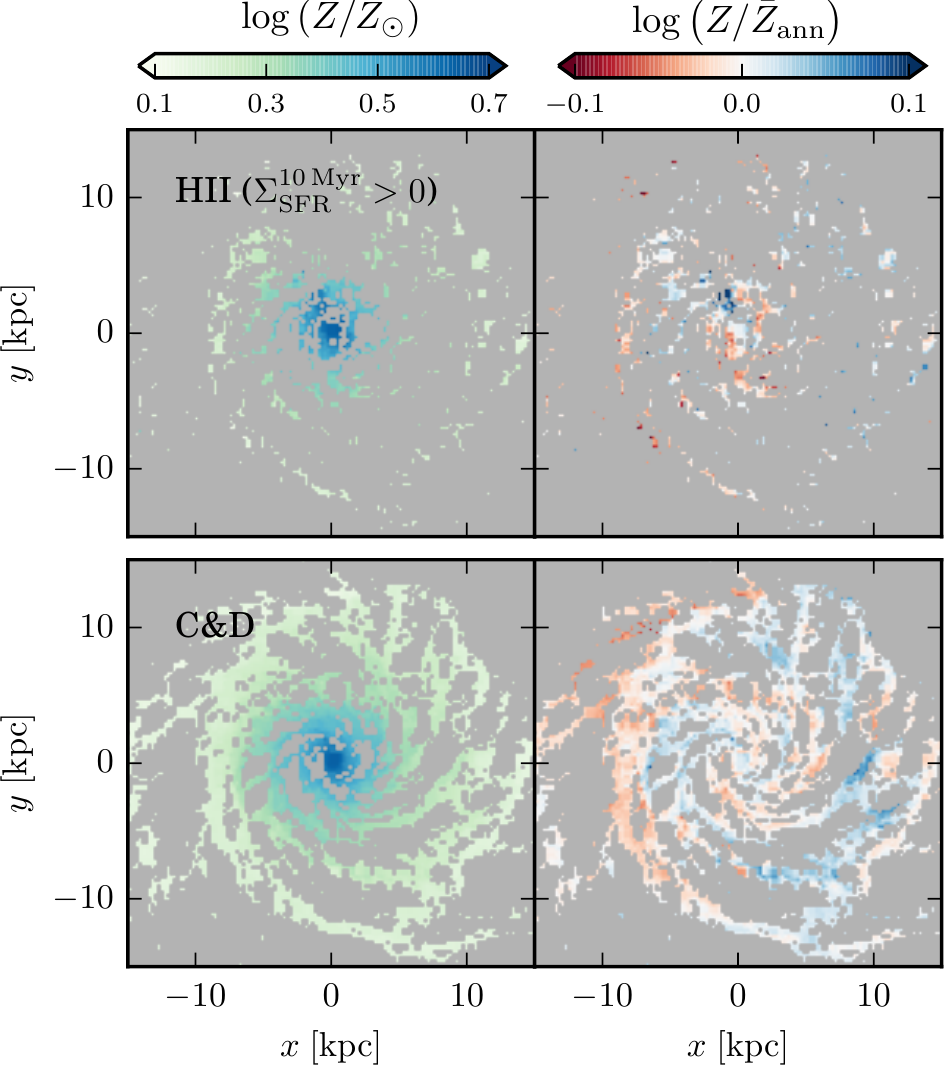}
    \caption{Spatial distribution of gas-phase (total including all species tracked in FIRE-2) metallicity in \textbf{m12b} at $z \approx 0$.  \emph{Left column}: gas metallicity scaled to Solar. \emph{Right column}: gas metallicity normalized by mean gas metallicity within $\pm$250~pc galactocentric radius of each pixel. \emph{Rows (ISM phases), top to bottom}: `nebular' ionized hydrogen gas with $T \approx 10^4$~K and recent star formation $\Sigma_{\rm SFR}^{\rm 10 \, Myr}>0$, and Cold \& Dense gas (C\&D) with $T<500$~K and a surface density $\Sigma_{\rm C\&D} >  10$ \msolt~\pcsqt. 
    Holes in the cold and dense ISM are seen, corresponding to a combination of interarm regions and super-shells driven by supernova feedback. 
    The galactic metallicity gradient and the relative sparseness of the nebular regions compared to the cold dense gas is easily seen in the left column. Similar features appear in both ISM phases, with a notable `barber pole' pattern evident, and a relatively metal-rich starburst region seen in the nebular HII near the galactic center.}
    \label{fig:m12b_250pc_diffGasZs}
\end{figure}

We investigate gas phase metal abundances across the disks of six Milky Way/Andromeda-mass spiral galaxies from the `standard physics' Latte suite of FIRE-2 simulations introduced in \citet{Wetzel2016} and \citet{Hopkins2018:fire}. 
A previous work, \citet{Orr2020}, studied the properties of spatially resolved gas surface densities, velocity dispersions, and SFRs in detail. This work makes use of 71 snapshots near $z\approx 0$ spaced $\sim$25 Myr in time (with a subset having a finer, $\Delta t \approx 2.2$ Myr, spacing) for each of the simulations.  
A brief summary of the $z \approx 0$ global properties of the galaxy simulations analyzed here is included in Table~\ref{table:galprops}.  

\begin{table*}\caption{Summary of $z\approx0$ properties of the FIRE-2 Milky Way-like galaxies used in this work. }\label{table:galprops} 
\begin{tabular}{lccccccc}
\hline
Name & $\log(\frac{M_\star}{{\rm M_\odot}})$ & $\log(\frac{M_{\rm gas}}{{\rm M_\odot}})$ & $\frac{R_{\star,1/2}}{\rm kpc}$& $\frac{R_{\rm gas,1/2}}{\rm kpc}$ & $\frac{v_c}{\rm km/s}$\textsuperscript{\textdagger} & $\frac{\bar Z_{\rm gas }}{Z_\odot}^*$ & $\left<\frac{d\log Z_{\rm gas}}{dR}\right> [{\rm dex/kpc}]^\ddagger$\\
 \hline
m12b  & 10.8 & 10.3 & 2.7 & 9.4 & 266  & 2.2 & -0.043 \\
m12c  & 10.7 & 10.3 & 3.4 & 8.6 & 232  & 2.1 & -0.041 \\
m12f   & 10.8 & 10.4 & 4.0 & 11.6 & 248 &1.9 & -0.038 \\
m12i   & 10.7 & 10.3 & 2.9 & 9.8 & 232 & 2.1 & -0.043 \\
m12m & 10.9 & 10.4 & 5.6 & 10.2 & 283 & 2.9 & -0.036 \\
m12r   & 10.2 & 10.0 & 4.7 & 9.9 & 156 & 1.1 & -0.019 \\
 \hline
\multicolumn{8}{l}{Note: all quantities measured within a 30~kpc cubic aperture, unless otherwise noted.}\\
\multicolumn{8}{l}{\textsuperscript{\textdagger}Circular velocities evaluated at $R_{\rm gas,1/2}$.}\\
\multicolumn{8}{l}{$^*$Average (mass-weighted) gas metallicity measured within $R_{\rm gas,1/2}$.}\\
\multicolumn{8}{l}{$^\ddagger$Least-squares fit for average (mass-weighted) gas metallicity gradient measured within $R_{\rm gas,1/2}$.} \\
\end{tabular}
\end{table*}

\begin{figure*}
	\includegraphics[width=0.97\textwidth]{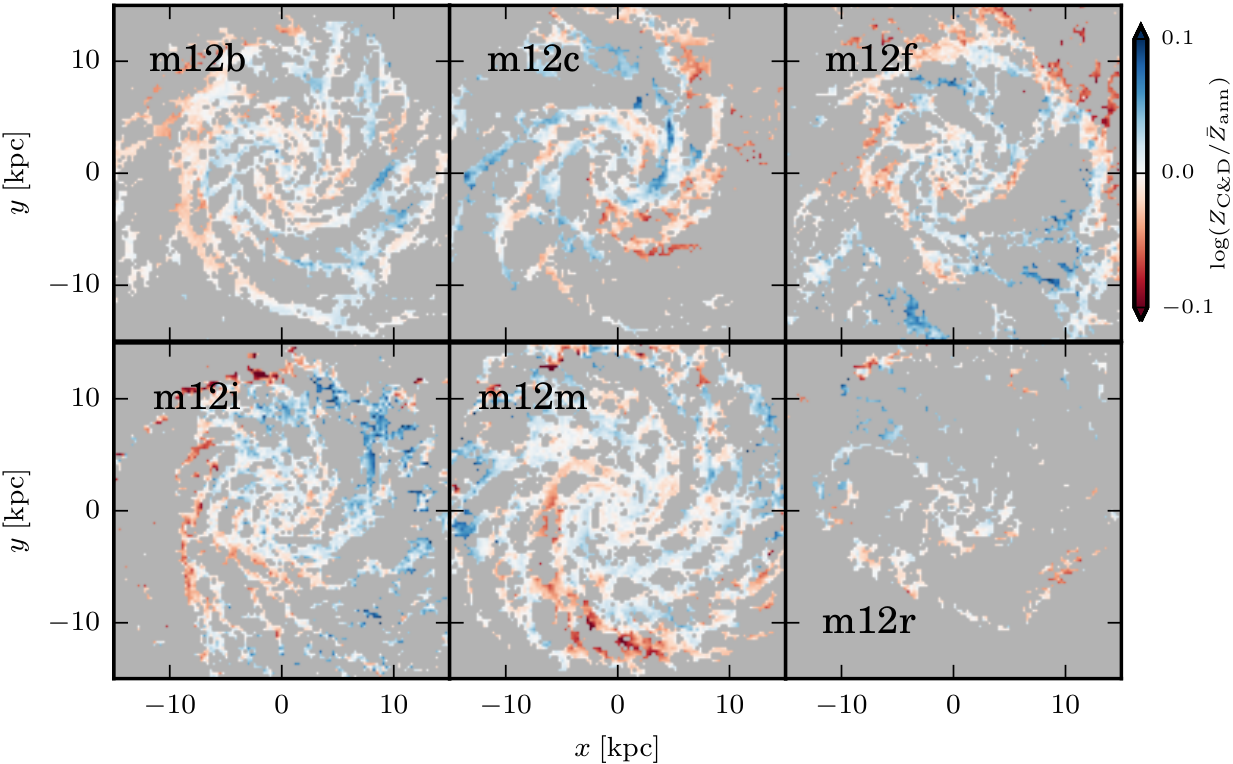}
    \caption{Cold dense gas-phase azimuthal metallicity variations, relative to mean cold dense gas metallicity within $\pm$250 pc at a given radius, in the six FIRE-2 galaxies at $z\approx 0$ with a 250 pc pixel size. Surface density cut such that $\Sigma_{\rm C\&D} > 10$ \msolt~\pcsqt, to approximate molecular gas tracers, was applied \emph{before} calculating ($R_{\rm gal}$$\pm$250~pc) mean metallicities. Grey regions denote areas with no data. Across the sample, relatively metal-rich and metal-poor regions appear to follow spiral arm structures and are coherent on kpc-scales. }
    \label{fig:Zcd_250}
\end{figure*}

\begin{figure*}
	\includegraphics[width=0.97\textwidth]{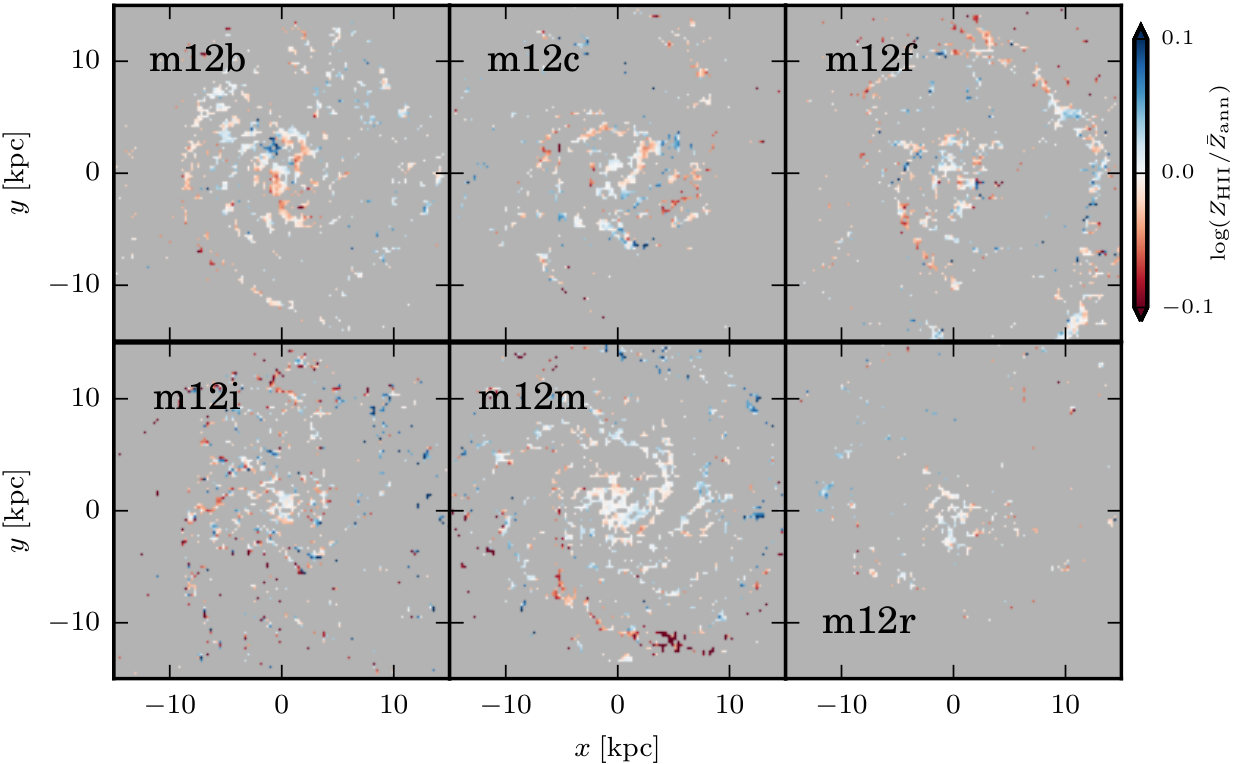}
    \caption{Nebular ionized hydrogen-phase ($T \approx 10^4$~K and $\Sigma_{\rm SFR}^{\rm 10\, Myr} > 0$) azimuthal metallicity variations, relative to mean nebular HII gas metallicity within $R_{\rm gal}$$\pm$250 pc at a given radius, in the six FIRE-2 galaxies at $z\approx 0$ with a 250 pc pixel size, as in Fig.~\ref{fig:Zcd_250}. Star formation rate cut was applied \emph{before} calculating annuli mean ionized hydrogen ($T \approx 10^4$~K) metallicities. Morphological and relative-metallicity structures seen in the cold dense gas are seen in the nebular gas, though with some differences: notably there are metal-rich star-bursting regions visible, and scattered relatively metal-poor regions of star formation that do not necessarily follow arm structures.}
    \label{fig:ZHII_250}
\end{figure*}

\begin{figure}
	\includegraphics[width=0.47\textwidth]{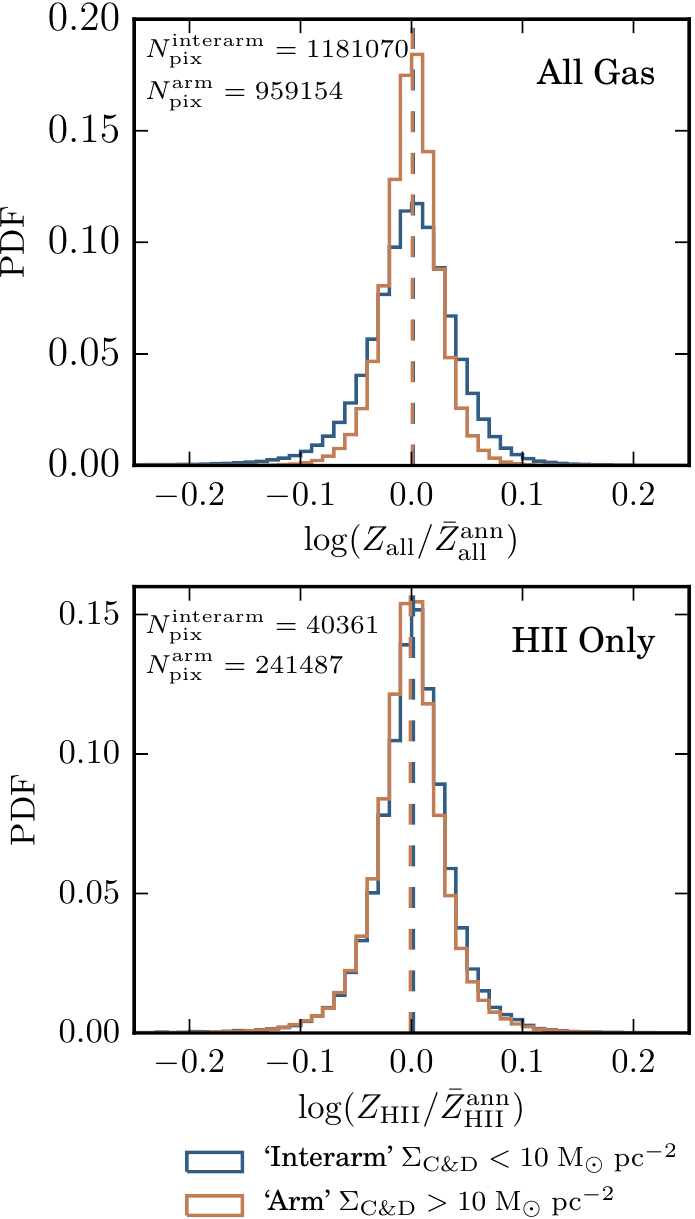}
    \caption{PDFs of gas-phase azimuthal metallicity variations in `Arm' and `Interarm' pixels, relative to mean gas metallicity within $\pm$250 pc at a given radius, in the six FIRE-2 galaxies at $z\approx 0$ with a 250 pc pixel size for all gas within $R_{\rm gas} < 10$~kpc. Dashed colored vertical lines indicate median of each distribution. For simplicity, we define `Arm' pixels with a surface density cut such that $\Sigma_{\rm C\&D} > 10$ \msolt~\pcsqt. We find no significant offset between the distribution of pixels in `arms' (\ie those dominated by cold dense gas) compared with `interarm' regions.  `Interarm' regions have a slightly wider distribution of azimuthal metallicity variations.}
    \label{fig:arm-interarm}
\end{figure}

The simulations analyzed here all have typical baryonic particle masses of $m_{\rm b,min} = 7100$ M$_\odot$, minimum adaptive force softening lengths $<$1~pc, and a 10~K gas temperature floor.  
With adaptive softening lengths, we note that the median softening length within the disk in the runs at $z=0$ is $h\sim 20-40$ pc (at a $n\sim 1$ cm$^{-3}$), with the dense turbulent disk structures having necessarily shorter softening lengths. 
The `pixel sizes' considered in this work are $250$ to $750$ pc, such that the minimum resolvable scales in the simulations always fit well within them (\eg when considering cold dense gas above HI-to-H$_2$ transition surface density, $\sim$10 \msolt \pcsqt, this corresponds to $\gtrsim$90 gas cells per 250 pc pixel), and are comparable with IFU observations \citep{Cortese2014, Bundy2015, Leroy2021}.

Star formation in the simulations occurs on a free-fall time in gas which is dense ($n >10^3$ cm$^{-3}$), molecular (per the \citealt{Krumholz2011} prescription), self-gravitating (viral parameter $\alpha_{\rm vir} < 1$) and Jeans-unstable below the resolution scale. Star particles are treated as single stellar populations, with known age, metallicity, and mass.  
Feedback from supernovae, stellar mass loss (OB/AGB-star winds), photoionization and photoelectric heating, and radiation pressure are explicitly modeled.  
These simulations do not include any supermassive black holes (SMBHs), and accordingly do not have any AGN feedback, nor do they include cosmic rays or other MHD physics (though simulations in the FIRE-2 suite have explored implementations of those `extended physics', \citealt{Angles-Alcazar2017, Chan2019, Su2019,  Angles-Alcazar2020}). 
Detailed descriptions of the `standard' physics and their implementation can be found in \citet{Hopkins2018:fire}.

Though \citet{Hopkins2018:fire} explicitly describes the implementation/abundance patterns of the FIRE metal yields, we summarize here for the reader: nucleosynthetic yields from core-collapse SNe are from \citet{Nomoto2006}, Type-Ia SNe yields are derived from \citet{Iwamoto1999}, and stellar wind yields (from O, B, and AGB stars) are from a set of models compiled in \citet{Wiersma2009}. The core-collapse and Type-Ia SN rates, respectively, are adopted from {\scriptsize STARBURST99} \citep{Leitherer1999} and \citet{Mannucci2006}.

Of particular importance for this study, the simulations investigated here also include a sub-grid metal diffusion/mixing model between gas cells, presumably occurring in unresolved turbulent eddies \citep[][see the latter for a more detailed description of the method's implementation]{Su2017, Escala2018, Hopkins2018:fire}.
Sub-grid modelling the turbulent diffusion of metals is necessary in  galaxy-scale simulations because of the ubiquity of turbulence in realistic environments \citep{Burkhart2010,Pingel2018,Mocz2018,Burkhart2021}. 
In FIRE-2, the turbulent metal diffusion term smooths the abundance following the prescription of \citet{Shen2010} based on a model from \citet{Smagorinsky1963}, and results in a more realistic distribution of gas phase metallicities. 
It also produces a more realistic stellar metallicity distribution \citep[][]{Escala2018}. 
To reiterate the diffusion implementation explicitly here:
\be\label{eq:diffusion}
\begin{split}
\frac{\partial M_i}{\partial t} + \nabla \cdot (D \nabla M_i) = 0 \; , \\ 
D = C_0 \parallel\textbf{S}\parallel_f h^2 \; ,
\end{split}
\ee
where $h$ is the smoothing length for a given gas cell, $C_0$ is a constant (proportional to the Smagorinsky-Lilly constant) calibrated with simulations by \citet{Su2017} and \citet{Hopkins2018:fire}, and $\textbf{S}$ is the symmetric traceless shear tensor defined as
\be 
\textbf{S} = \frac{1}{2}(\nabla\textbf{v} + (\nabla\textbf{v}^T))-\frac{1}{3}Tr(\nabla\textbf{v}) \; ,
\ee
with $\textbf{v}$ being the local shear velocity.  
The simulations analyzed here have been run with $C_0 = 0.008$ as the default parameter. We investigate the dependence of the large-scale metallicity transport/azimuthal abundance variations seen here on the strength of the mixing/diffusion coefficient $C_0$ in \S~\ref{sec:tmd}.

We produce mock observational maps from the snapshots using the same methods as \citet{Orr2018} and \citet{Orr2020}, projecting the galaxies face-on according to the angular momentum of the star particles within the stellar half-mass radius, and binning star particles and gas cells into square pixels with side-lengths (\emph{i.e.}, ``pixel sizes'') 250-750~pc. The maps are 30~kpc on a side, and integrate gas and stars within $\pm 15$ kpc of the galactic mid-plane. 

We generate a proxy for observational measures of recent SFRs by calculating the 10 Myr-averaged SFR. We do this by summing the mass of star particles with ages less than 10 Myr, and correcting for mass loss from stellar winds and evolutionary effects using predictions from {\scriptsize STARBURST99} \citep{Leitherer1999}.  
This time interval was chosen for its \emph{approximate} correspondence with the timescales traced by recombination lines like H$\alpha$ \citep{Kennicutt2012}\footnote{Recent work by the late Jose Flores Velazquez \citep{Velazquez2020} indicates that H$\alpha$ may indeed trace star formation averages on timescales closer to 4 Myr.}, and to associate ionized gas near $T \approx 10^4$ K with star-forming regions as a proxy for identifying HII region nubulosity.  
For comparability with dense gas observations, we calculate column densities for the ``cold and dense'' gas ($\Sigma_{\rm C\&D}$ throughout) with $T < 500$~K and $n_{\rm H} > 1$ cm$^{-3}$.  
This gas reservoir taken as a proxy for the cold molecular gas in the simulations following the methodology of \citet{Orr2020}, and ought roughly to correspond with gas traced by cold dust or CO observations.  
As well, we calculate a column for the warm ionized medium (WIM), where we identify gas near the $T \approx 10^4$ K ridge-line (specifically, $|\log T - 4.05 | < 1/6$).  
Throughout, we identify this gas reservoir with warm ionized gas as constituting `HII regions' in star-forming clouds (when cospatial with the recent 10 Myr star formation tracer).  
We suggest the reader compare our `HII region' WIM columns (and associated metallicities) with nebular emission observations of \eg [OIII].

\section{Results}
Figure~\ref{fig:m12b_250pc_diffGasZs} shows the resulting spatial metallicity distributions for one of the simulated galaxies as an example of the maps produced in our analysis, \textbf{m12b}, at $z \approx 0$ in the two phases of the ISM: the ionized nebular HII gas near $T \sim 10^4$~K (where $\Sigma_{\rm SFR}^{\rm 10 \; Myr} > 0$) and the cold \& dense (C\&D; $T < $~500~K).  Figure~\ref{fig:m12b_250pc_diffGasZs} also shows the resulting `barber pole' pattern when we subtract the mass-weighted mean metallicity of all pixels with galactocentric radii $\pm$250 pc for each pixel.  Subtracting the `local mean' at each pixel's galactocentric radius essentially subtracts the radial metallicity profile non-parametrically, allowing us to explore the azimuthal variations.
This reveals an alternating pattern, evident in both ISM phases, of metal-rich and metal-poor gas filaments leading from the core of the galaxy to the edge of the gas disk.  
The overall gas-phase metallicity gradients \citep[explored in-depth for these simulations by][]{Bellardini2021} are consistent between ISM phases (\ie the metal reservoir is well-distributed and mixed between cold, cool and ionized gas).  
Holes, consistent with inter-arm regions and supernova-drive super-bubbles, are evident in the cold and dense gas phase. 
In the following analysis, we will focus our attention on the mean metallicity-subtracted pixel distributions, to understand origins of the azimuthal metallicity variations seen. 

\subsection{Structure in Azimuthal Metallicity Variations: Spiral Arms}\label{sec:AzZstructure}
Figures~\ref{fig:Zcd_250} and \ref{fig:ZHII_250} show metallicity variations, relative to the mass-weighted azimuthal averages, in the cold dense and nebular HII gas, respectively.  
In analyzing the local (pixel-scale) variations, we compare local abundances with the mass-weighted abundance of pixels within a $\pm$250 pc galactocentric radius band, after having taken the appropriate density/ionization/temperature/star-formation rate cuts.  
Thus, we are essentially comparing local abundances to the (mass-weighted mean) metallicity gradient, with 500~pc (wide top-hat function) radial smoothing.  
We explore whether there is any apparent difference, spatially, in comparing total gas-phase metallicity, Oxygen, or Iron abundances in cold dense gas in Appendix~\ref{sec:appendix:ZOFe}.  We find no significant difference in the various abundance distributions, and so use total metallicity (including all abundances that are tracked in FIRE-2) for our analysis throughout the remainder of the main text.

Intriguingly, the pattern of variations across all of the simulations (with the possible exception of \textbf{m12r}, which has a much less massive, sparser gas disk than the others, owing to recent interactions with multiple LMC-mass companions) coherently follows the flocculent spiral arms from the galactic centers to their outskirts in a `barber shop pole'-like fashion of alternating metal-enriched (blue) and metal-poor (red) arms.  
Largely, the metallicity variations in the two phases appear to track one another, though the nebular regions are spatially scattered and do not cover a large fraction of the disks. 

We explore the degree to which the distributions of azimuthal metallicity variations are different between the arm and interarm regions (defined with a very simple cold dense gas surface density cut of $\Sigma_{\rm C\&D} > 10$ M$_\odot$ pc$^{-2}$) for the total gas column (no cuts) and the HII nebular regions in Figure \ref{fig:arm-interarm}. We find no significant bias between the metallicity variance distributions, in either the total or HII nebular gas phases, indicating strongly that any offsets in azimuthal metallicity variance is not driven significantly by arm-interarm differences.  In the total gas reservoir, there is a slightly wider distribution of azimuthal metallicity variations in the interarm pixels.  However, this can be attributed to the fact that the interarm pixels have on-average slightly lower dense gas turbulent velocity dispersions, and for the same length (pixel) scale, this corresponds to slightly longer turbulent crossing/mixing timescales.  And so, a slightly wider distribution in metallicity variations ought to be expected in the interam regions, even with no distinguishable offset in the distribution from the arm regions.

\subsection{Metallicity Variations between ISM Phases are Small}
\begin{figure}
	\includegraphics[width=0.47\textwidth]{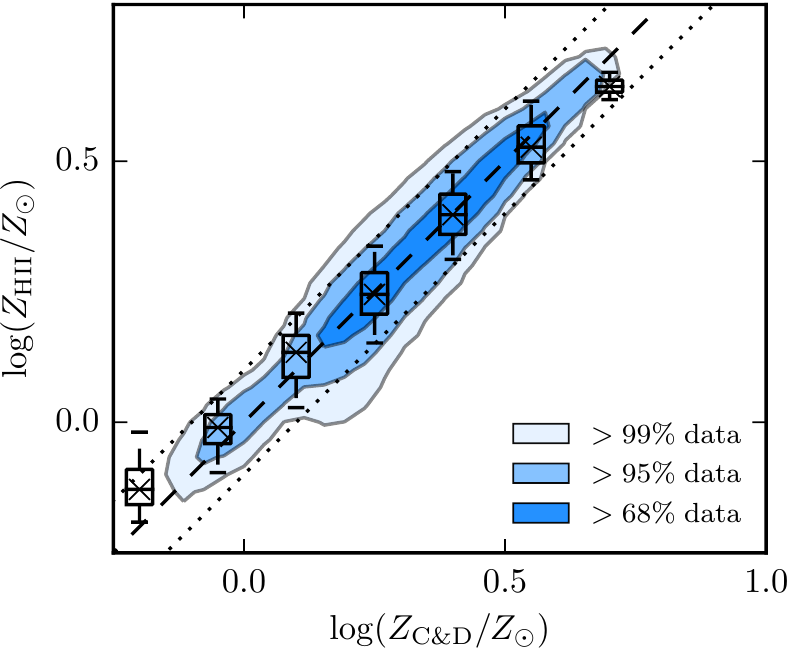}
    \caption{Metallicity (scaled to Solar) in cold and dense gas ($T< 500$ K, $\Sigma_{\rm C \& D} > 10$ \msolt~\pcsqt) compared to HII nebular regions ($T \approx 10^4$ K, $\Sigma_{\rm SFR}^{\rm 10\,Myr} > 0$) in the six the FIRE-2 galaxies in 250~pc pixels for all snapshots. Dashed and dotted lines indicate unity and $\pm 0.1$ dex, respectively. Colored contours indicate data inclusion regions of 68, 95, and 99\%.  Bar and whiskers indicate interquartile, 5-95\% range, and median values in 0.15~dex wide bins of $Z_{\rm C\&D}$.  In the overwhelming majority of pixels, the metallicity of nebular regions and cold \& dense gas is within $\pm 0.1$~dex of each other.}
    \label{fig:Zcd_vs_ZHII_250}
\end{figure}

\begin{figure}
	\includegraphics[width=0.47\textwidth]{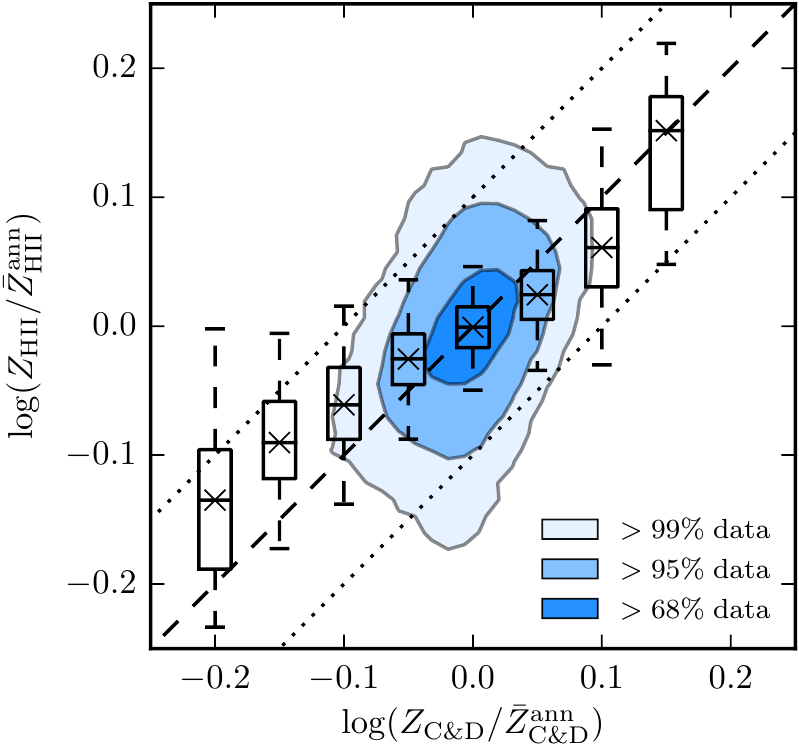}
    \caption{Variance relative to azimuthal mean metallicity (for each respective gas phase, 0.5~kpc-wide annuli) in cold and dense gas compared to HII nebular regions, style and data selection as in Figure~\ref{fig:Zcd_vs_ZHII_250} (box and whisker bins are 0.05 dex wide here).  Metallicity variations in cold and dense gas are somewhat smaller than in nebular regions, driven perhaps by shorter mixing timescales. At `very' low relative-metallicity ($<-0.1$ dex) in cold and dense gas, nebular regions are relatively metal-rich, suggesting local enrichment contributions.  The metallicity variations are well correlated between the two gas phases on-average, though variances in nebular regions are more extreme than in cold dense gas within the 95\% data inclusion envelope. }
    \label{fig:varZcd_vs_varZHII_250}
\end{figure}

An observational difficulty in connecting metallicity measurements and disk dynamics is that the most common gas-phase metallicity tracer is nebular emission from [OIII] whereas the kinematics of the dense gas of arm structures is often traced by sub-mm emission such as the rotational transitions of CO. We might be concerned that the overall metallicity and azimuthal metallicity variations seen in nebular HII do not trace that in the cold dense gas.  Figures~\ref{fig:Zcd_vs_ZHII_250} and \ref{fig:varZcd_vs_varZHII_250} address this directly.

Figure~\ref{fig:Zcd_vs_ZHII_250} demonstrates that on 250 pc-scales, it is overwhelmingly the case (95\% of the pixels in our dataset) that the overall metallicity is equal within $\pm 0.1$~dex between the two ISM phases. This is on the order of the overall metallicity scatter, though it should be noted that there is no systematic offset of the metallicities between the ISM phases.  Interpreted differently, the radial metallicity profiles in nebular HII and cold dense gas appear to be the same, albeit with $\sim$0.1 dex scatter locally (pixel-to-pixel) relative to each other.  We note that the lobe of data near Solar metallicity primarily comes from \textbf{m12r} in this plot, as its overall metallicity normalization is $\sim$0.5~dex lower than the other five Milky Way mass simulations. 

In Figure \ref{fig:varZcd_vs_varZHII_250}, we explore how the azimuthal metallicity variations, rather than the overall metallicity normalization, in the nebular HII and the cold dense gas correlate on a pixel-to-pixel basis with a 250~pc pixel size.  Restated, are pixels that are metal-rich in the HII nebular tracer also metal rich in the cold dense tracer, compared to their like surroundings (and \textit{vice versa} with respect to metal-poor pixels)? To find the variations, we first calculate the annular mass-averaged metallicity in each phase for each pixel considering all the pixels within $\pm$250 pc of its galactocentric radius.
The relative scatter in nebular HII metallicities is larger than the cold dense gas in the 2-3$\sigma$ tails of the distribution between the 68-99\% contours, which likely connects to the shorter mixing timescale of cold dense gas (it is more highly supersonic than the WIM).  
At the `extreme' metal-poor end of the metallicity variations in cold dense gas ($\log(Z_{\rm C\&D}/\bar Z_{\rm ann}) < -0.1$), there appears to be a bias toward less metal-poor variation in nebular HII, which is likely indicative of the effects of local enrichment/star formation.  
Generally, the azimuthal metallicity variations in both the nebular HII regions and cold dense gas are well-correlated with the core of the pixel distribution lying within $0.05$~dex of equality.

\subsection{Lack of Correlation with Star Formation Surface Density}\label{sec:noSFRcorrelation}
\begin{figure*}
	\includegraphics[width=0.97\textwidth]{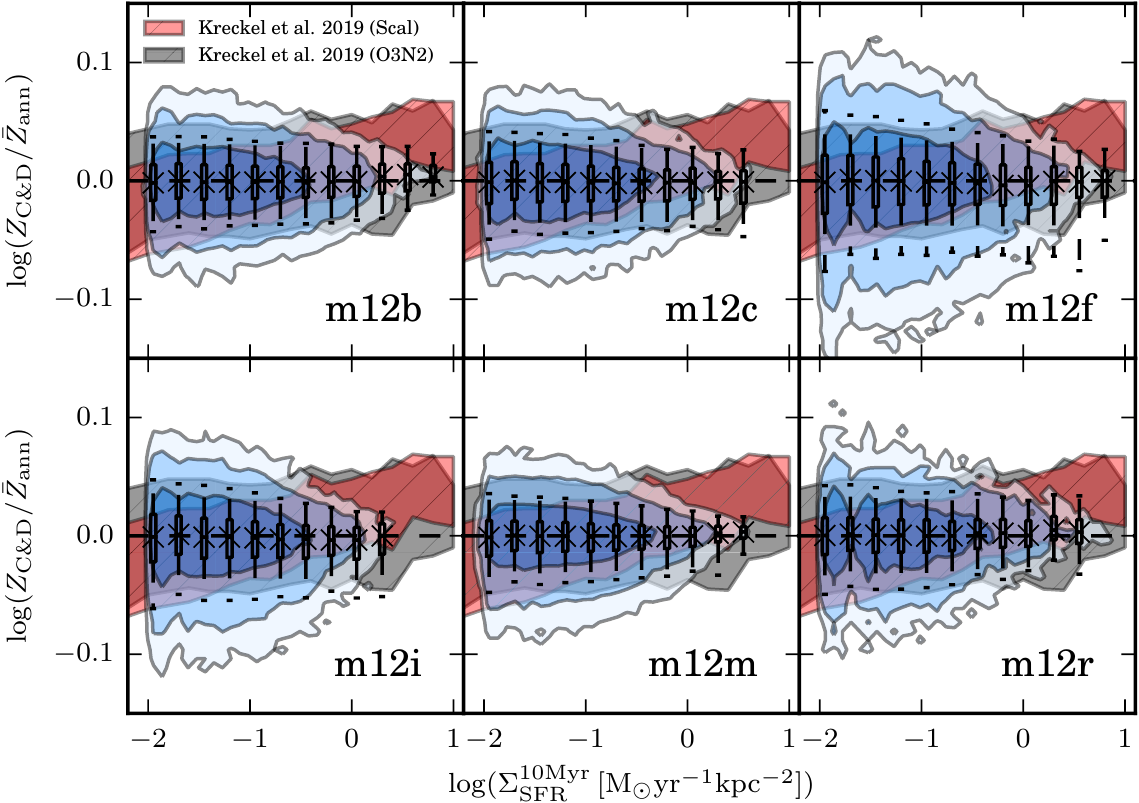}
    \caption{Distribution of metallicity relative to median median metallicity (within a galactocentric radius of $\pm$250 pc for each pixel) as a function of 10~Myr-averaged SFR in six FIRE-2 spiral galaxies.  Blue contours in each panel indicate 50\%, 75\%, and 95\% data inclusion regions. Bars and whiskers show 5-95\% range, and inter-quartile regions (along with average value marked with an `x'), in 0.25~dex-wide SFR bins.  Red and grey filled contours indicate data range from \citet[][combined with \citealt{Kennicutt2012} H$\alpha$--$\dot M_\star$ calibration]{Kreckel2019}, using two metallicity calibrations, of gas-phase azimuthal variations in local spiral galaxy HII regions. Though the contours are biased towards the outskirts of the galaxies (there are more pixels at larger radii), there is good agreement between the full distribution of data and the data binned in SFR.  No significant trend in azimuthal metallicity variation is seen as a function of SFR, pixels at constant SFR appear as likely to be relatively metal enriched or poor.  Scatter in metallicity appears to be somewhat weaker at high SFR, though it is unclear if this is related more to shorter dynamical times in galaxy centers.  \emph{Local star formation does not appear to be directly correlated with the coherent variations in metallicity observed in these flocculent spirals.}}
    \label{fig:varZcool_SFR10_100}
\end{figure*}

Figure~\ref{fig:varZcool_SFR10_100} shows the distribution of metallicity, relative to the mass-weighted mean metallicity at each pixel's galactocentric radius $R_{\rm gal}$ $\pm$250 pc, in the cold dense gas, as a function of the local 10 Myr-average SFR surface density.
Both in terms of the overall distribution, and as a function of SFR, there appears to be no significant trend in average relative metallicity with SFR. 
Though the scatter in relative metallicity is diminished somewhat at high SFR surface density, this is likely due to the combined geometric effect of outskirts having more pixels and that the average SFR surface density is strongly radially dependent, with galaxy centers having shorter dynamical times (and thus being more effectively mixed). Similarly, the 100 Myr-average SFR (not plotted) shows broadly the same qualitative result, though at very high SFRs there is a very slight bias towards higher relative metallicities ($\sim$0.01~dex), in-line with sustained self-enrichment. \citet{Bellardini2021} also quantified the degree of azimuthal scatter associated with considering partial azimuthal arcs, with lengths from $\sim$500~pc to the full annulus, finding only modest rises in the degree of scatter from $\sim$0.05 to $\sim$0.08~dex. The azimuthal variation in gas-phase metallicity is consistent with not being driven by local enrichment from star formation, pointing at another, dynamical in origin, cause for the $\sim$0.1~dex scatter.   

Indeed a back of the envelope calculation might suggest that the self-enrichment of gas following a star formation event due to a spiral arm passage is smaller than the observed scatter.  
Taking a spiral arm to be modeled as a massive GMC with $M_{\rm GMC} \sim 10^6$ \msolt and Solar metallicity, we might expect a star cluster of approximately $\sim$10$^4$ \msolt~(a typical star formation efficiency of $\sim$1\%) to form.  
Integrating down the IMF, estimating the number of core-collapse SNe and their ejecta, we might expect the star cluster to return $\sim$130 \msolt~of new metals to the ISM (\citealt{Agertz2013} explicitly lays out this calculation in their modeling of feedback from a star cluster).  
And so in this case, the variation that we might expect to see ought to be of the order $\Delta\log Z \sim \log((M_{\rm GMC}Z_\odot+M_{\rm Z, new})/M_{\rm GMC})-\log(Z_\odot) \sim \log(1+ \frac{M_{\rm Z, new}}{M_{\rm GMC}Z_\odot}) \sim 0.003$~dex. 
For comparison, the scale of the azimuthal variations seen is more on the order $\pm 0.025-0.05$~dex, larger by a factor of a few to an order of magnitude than this estimate.
And as a cluster mass of $\sim$10$^4$ \msolt~is reasonably consistent with the star formation rate occurring on the 250~pc-scales that we are analyzing ($\Sigma_{\rm SFR}^{\rm 10\, Myr} l_{\rm pix} \Delta t  \approx 0.1 \; {\rm M_\odot \; kpc^{-2} \; yr^{-1}}\times (0.25 \; {\rm kpc})^2  \times 10^7 \; {\rm yr} \approx 10^4 \; {\rm M_\odot}$), it is not entirely surprising that any given spiral arm passage/star formation event does not result in significantly elevated metallicity, relative to the annular average.  
Further, since the gas abundance in these simulations at $z\approx 0$ is super-Solar, the $\Delta \log Z$ from a single star cluster is likely even smaller.  

We compare our data against spatially resolved observations of HII regions in nearby spirals by \citet{Kreckel2019} taken as part of the PHANGS-MUSE survey.  We take their data of azimuthal gas-phase [O/H] variations, spatially resolved to $\sim$50~pc, as a function of H$\alpha$ luminosity, and convert it to a SFR surface density estimate using the \citet{Kennicutt2012} $L_{\rm H\alpha}$--$\dot M_\star$ calibration and their approximate spatial resolution of 50~pc.  The relative degree of scatter as a function of SFR is in good agreement between the simulated and observed galaxies, across $\sim$3~dex in SFR.  However, \citet{Kreckel2019} found a slight positive correlation with H$\alpha$ luminosity (SFR) when using their `Scal' calibration, whereas essentially a flat relation using the \citet{Marino2013} O3N2 calibration.  The slight positive correlation is in tension with our high-SFR regions, for $\sim\Sigma_{\rm SFR} > 1$ \msolt~yr$^{-1}$ kpc$^{-2}$. Other work by \citet{Grasha2019} has found a flat relation of gas-phase azimuthal metallicity variations with ISM pressure, and to the extent that ISM pressure scales linearly (or nearly so) with local SFR \citep{Shetty2012, Faucher-Giguere2013, Kim2013, Kim2015c, Gallagher2018, Fisher2019, Orr2019, Ostriker2022}, this corroborates a flat or slightly positive dependence of azimuthal gas-phase variations with SFR.

\subsection{Gas Flows Along Spiral Arms as Source of Azimuthal Metallicity Variations} \label{sec:gasflow}
\begin{figure*}
	\includegraphics[width=0.97\textwidth]{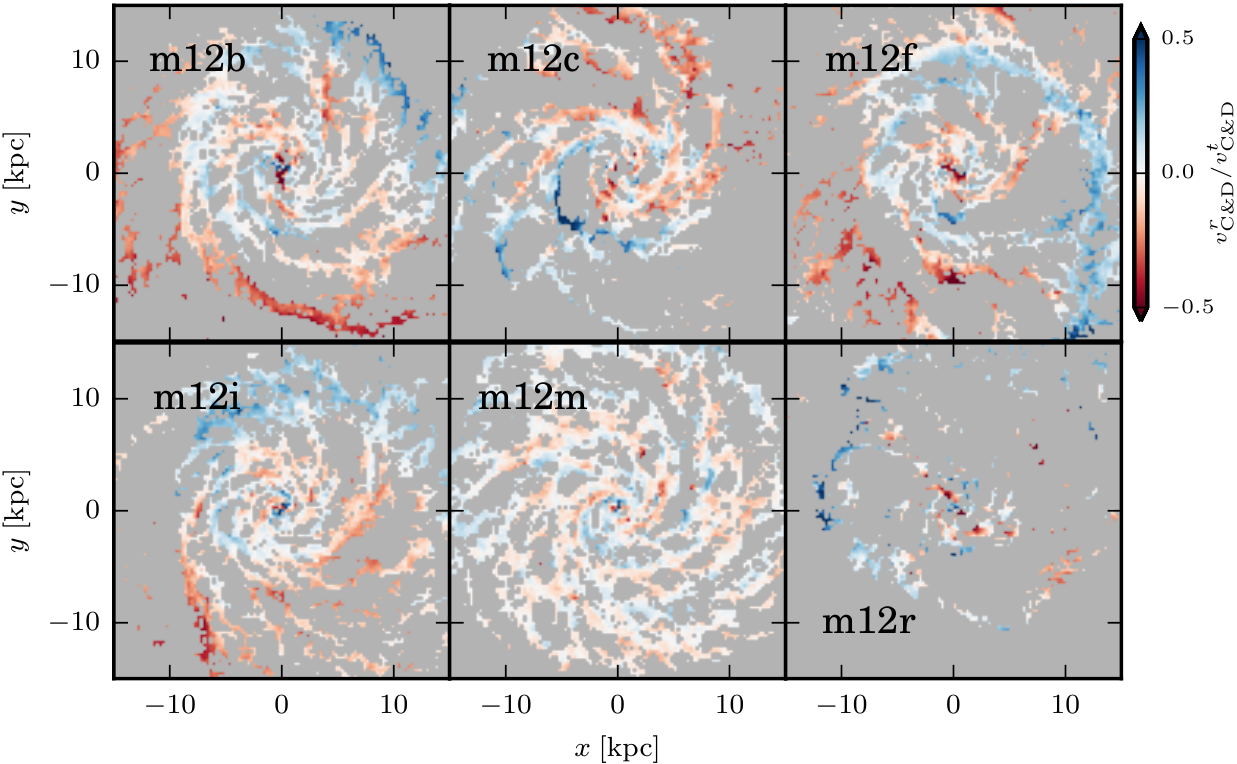}
    \caption{Six FIRE-2 Spiral Galaxies at $z\approx 0$ colored by local ratio of radial to tangential gas velocity averaged over 100 pc scales (pixels), with blue/red colors representing outwardly/inwardly moving gas.  Spiral arm structures in these flocculent spirals have coherent net inward or outward motion, appearing to alternate from one arm to the next. A marked similarity can be noted with the patterns seen in Figure~\ref{fig:Zcd_250}, suggesting the connection between azimuthal metallicity variations and large-scale velocity structures.  A handful of expanding holes in the ISM are also seen, notably in \textbf{m12c}.}
    \label{fig:Vcd_250}
\end{figure*}

\begin{figure*}
	\includegraphics[width=0.97\textwidth]{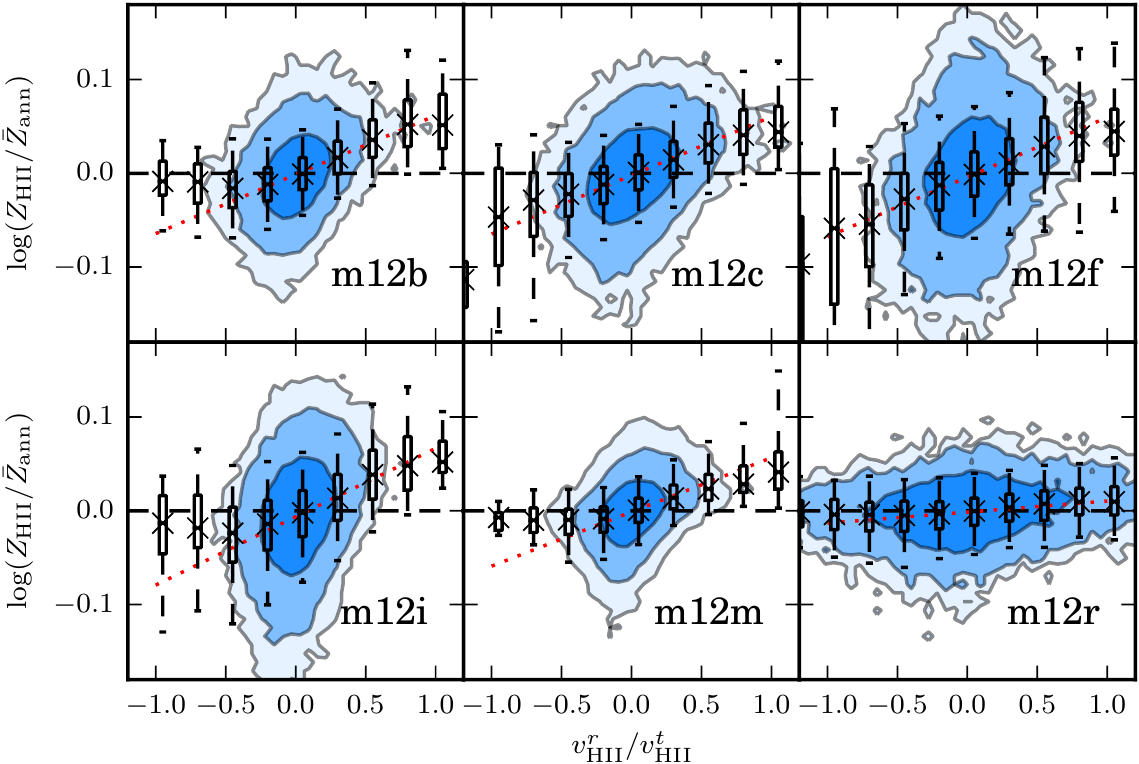}
    \caption{Distribution of metallicity relative to median median metallicity in HII nebular regions (within a galactocentric radius of $\pm$250 pc for each pixel) as a function of radial to tangential gas velocity ratio in six FIRE-2 spiral galaxies.  Blue contours and bar and whiskers in the style of Figure~\ref{fig:varZcool_SFR10_100}. Red dotted line indicates best linear least-squares fit to all pixels. Blue contours show that the majority of gas parcels have small positive/negative ratios of radial to tangential velocity, but that those parcels span much of the range of the azimuthal metallicity variations.  Binning the data by radial to tangential velocity ratio reveals a fairly coherent pattern of outwardly moving relatively metal enriched gas, and inwardly moving relatively metal poor gas.  This pattern is present in all six galaxies to some degree, suggesting that large-scale coherent inward/outward motions are responsible for much of the azimuthal metallicity variations in these flocculent galaxies.}
    \label{fig:varZHII_VHII_250}
\end{figure*}

\begin{figure*}
	\includegraphics[width=0.97\textwidth]{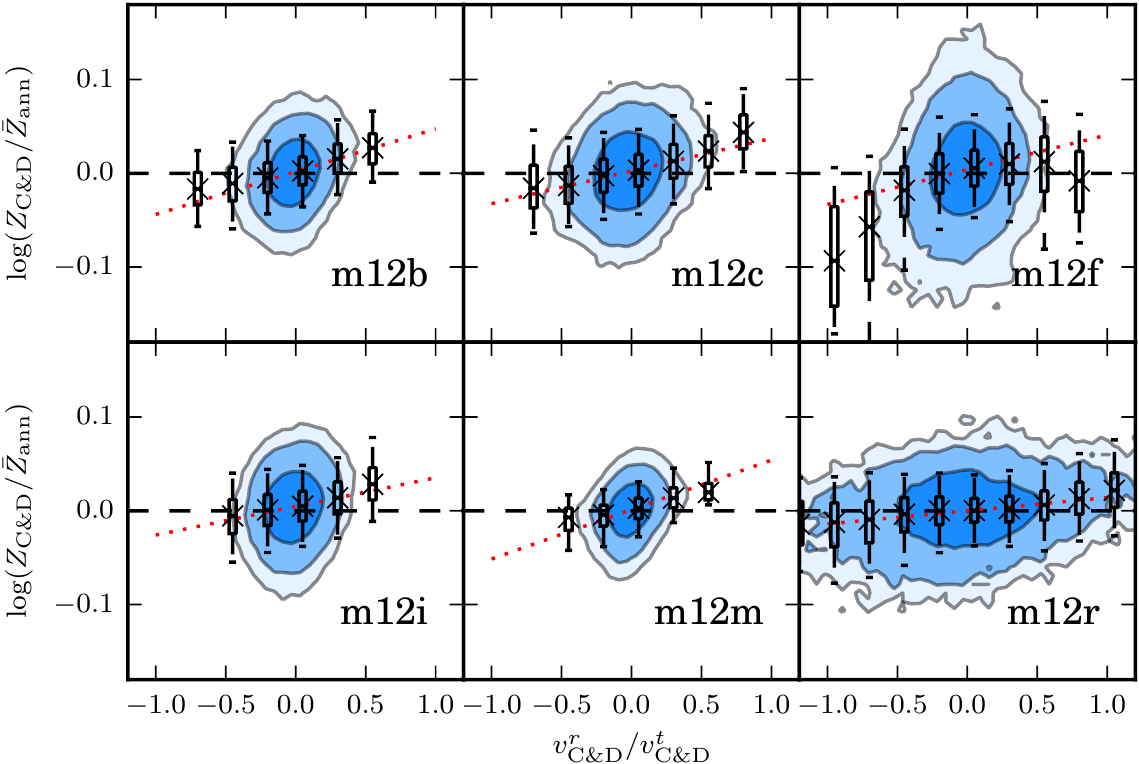}
    \caption{Distribution of metallicity relative to median median metallicity in cold dense gas (within a galactogentric radius of $\pm$250 pc for each pixel) as a function of radial to tangential gas velocity ratio in six FIRE-2 spiral galaxies.  Blue contours, bars and whiskers, red dotted line in the style of Figure~\ref{fig:varZHII_VHII_250}. Blue contours show that the majority of gas parcels have small positive/negative ratios of radial to tangential velocity, but that those parcels span much of the range of the azimuthal metallicity variations.  Binning the data by radial to tangential velocity ratio reveals a fairly coherent pattern of outwardly moving relatively metal enriched gas, and inwardly moving relatively metal poor gas.  This pattern is present for all six galaxies to some degree, suggesting that large-scale coherent inward/outward motions are responsible for much of the azimuthal metallicity variations in these flocculent galaxies.}
    \label{fig:varZcd_Vcd_250}
\end{figure*}

If local self-enrichment from recent star formation does not appear to be the primary source of the azimuthal metallicity variations, then we might search for kinematic or dynamical sources in the gas.  Figure~\ref{fig:Vcd_250} shows the six galaxies colored by the ratio of radial to tangential velocity ($v^r/v^t$) in the cold dense gas, averaged over 250~pc scales.  
Suggestive of a source for large-scale variations in gas properties, we can immediately see that there are large-scale coherent structures in radial velocity in the cold dense gas.  
Though not plotted, generally speaking, the nebular HII regions are embedded in the spiral structure and share a very similar velocity field. We see a similar pattern of alternating structures in the arms.  Especially in the most-flocculent cases of \textbf{m12b}, \textbf{m12i} and \textbf{m12m}, the arm structures appear to alternate between moving inwards and outwards relative to each other as part of the disk-scale shearing motions.  
These arm motions are consistent between snapshots across the whole of the analysis period ($\sim$1.4 Gyr), more broadly illustrative of the arms acting as freeways transporting gas inwards and outwards radially over several galactic dynamical times.

Quantifying the relationship between the azimuthal metallicity variations and the large-scale gas flows, Figures~\ref{fig:varZHII_VHII_250} \& \ref{fig:varZcd_Vcd_250} compare the average radial to tangential velocity ratios of the pixel distributions for all the snapshots against their azimuthal relative metallicities in the two ISM gas phases.  
Over-plotting a least-squares fit, we can see that there is a consistent positive relationship between azimuthal metallicity variations and radial velocity in both the nebular HII and the cold dense gas components in at least five of the simulations (the relation appears to be somewhat marginally `detected' in \textbf{m12r}, however it is the most disrupted and least-disky of the sample, provide somewhat of a `control' to test the transport model against).  We report the least-squares fits for the slopes $\eta$ in Table~\ref{table:eta_fits}, where we have fit a linear function $\log(Z/\bar Z_{\rm ann}) = \eta (v^r/v^t)+ \delta$ for each ISM phase at 250~pc and 750~pc pixel sizes.
It is also clear that the scatter in the azimuthal metallicity variations is larger in degree than the slope of the relation to the velocity, \ie even subtracting out the mean slope would result in significant remaining azimuthal metallicity variations.  
By eye, and in comparing values in Table~\ref{table:eta_fits}, the slopes of the metallicity variations versus radial velocities are greater in the nebular HII gas as compared to the cold dense phase.  This is likely related to the shorter mixing timescales in the highly turbulent cold dense gas, and the spatially more separated distribution of HII gas.

We find that there is a very weak decrease in azimuthal metallicity scatter with increased velocity dispersion in the cold dense gas, suggesting that though mixing is clearly important in the ISM, these variations (or lack thereof) are not primarily driven by it.

\begin{table}\caption{Least-squares fit $\eta$, in $\log(Z/\bar Z_{\rm ann}) = \eta (v^r/v^t)+ \delta$ for each ISM phase at 250~pc and 750~pc pixel sizes in five of the FIRE-2 Milky Way-like galaxies used in this work.  m12r has been separated given its unique morphology.}
\label{table:eta_fits}
\centering
\begin{tabular}{rcccc}
		& \multicolumn{2}{c|}{250 pc}		& \multicolumn{2}{c}{750 pc}                \\ \hline
Name    & $\eta_{\rm HII}$ & $\eta_{\rm C\& D}$ & $\eta_{\rm HII}$ & $\eta_{\rm C\& D}$ \\ \hline
m12b 	& 0.063	 & 0.046 & 0.072 & 0.036 \\
m12c 	& 0.063  & 0.035 & 0.074 & 0.027 \\
m12f 	& 0.064  & 0.037 & 0.073 & 0.030 \\
m12i 	& 0.073  & 0.031 & 0.085 & 0.020 \\
m12m 	& 0.058  & 0.053 & 0.075 & 0.045 \\ \hline
m12r 	& 0.012  & 0.014 & 0.012 & 0.013 \\ \hline
\end{tabular}
\end{table}

\subsubsection{Simple Model for Metallicity Variations arising from Radial Gas Flows}

Presuming that the large-scale shear motions are the cause of the coherent azimuthal metallicity variations, we can construct a simple model predicting the magnitude of the variance.  
Let us ignore the involved process that gives rise to metallicity gradients in galaxies, including inside-out growth, gas accretion, evolution in wind metal loading, \textit{etc.}~\citep[see, \eg][]{Sharda2021}, and take there to be an underlying gradient $d\log Z/dr$.  
Given a metallicity gradient, we can assume that a parcel of gas carries its metals with it a radial distance $\gamma v_r t_{\rm eddy}$, where $v_r$ is the net radial velocity, $t_{\rm eddy}$ is the local turbulent eddy turnover time, and $\gamma$ is an order unity pre-factor describing the effective speed of metal mixing on the (largest) eddy scale, ultimately relating to local disk geometry/velocity patterns and/or local metal diffusion/mixing.  
However, we can relate the disk scale height (necessarily the size of largest coherent eddy) to the local velocity dispersion as $l_{\rm eddy} \approx H \approx \sigma/\Omega$.  
And so, the eddy turnover time is then $t_{\rm eddy} \approx H/\sigma \approx R/v_c$. Together, we might then expect a scaling
\be \label{eq:model}
\Delta \log Z \approx \gamma R  \frac{ v_r}{v_c} \frac{{\rm d}\log Z}{{\rm d}r}\Bigg|_{r=R} \; .
\ee
Two reasonable limits for mixing/transport coincide with $\gamma = 0$ and $1$.  $\gamma = 0$ would imply perfect mixing, immediately smoothing out any azimuthal variations as the gas parcel moves in or outwards.  
On the other hand, mixing only on the largest-eddy scale would yield $\gamma = 1$, as gas would carry its metal abundance inwards or outwards for one eddy turnover time.  
And so we reasonably expect that $0 < \gamma < 1$ for most realistic mixing scenarios in galaxy disks.

We estimate $\gamma$ for the each simulation and ISM phase on two spatial scales, as the particular morphology, magnitude of radial gas transport, and metallicity gradient is unique to that galaxy (we do however, stack the snapshots from each simulation).  
To calculate the metallicity gradients themselves, we fit them self-consistently using the same annuli that we used to calculate the mass-weighted mean metallicities $\bar Z_{\rm ann}$. \citet{Bellardini2021} explored various properties of the metallicity gradients themselves in these simulations across redshift.  Table~\ref{table:gamma_fits} shows the best-fit $\gamma$ for each galaxy, in each of the two ISM phases we explore, at two spatial scales (250~pc and 750~pc), which results in zero-residual slope for the data in Figures~\ref{fig:varZHII_VHII_250} \& \ref{fig:varZcd_Vcd_250}.  
We exclude \textbf{m12r} on the basis that it does not clearly exhibit a flocculent spiral arm disk, but instead exhibits a rather dynamically disrupted structure (the best-fit lines in Figures~\ref{fig:varZHII_VHII_250} \& \ref{fig:varZcd_Vcd_250} show a much wider distribution in $v^r/v^t$ and much-flatter relations for the metallicity variations).  
We find that the ``effective'' mixing time for nebular HII is longer than the cold dense gas (\ie $\gamma_{\rm HII} > \gamma_{\rm C\&D}$), likely due to the larger spatial separation between nebular regions.  
Moreover, on larger scales a steeper relation is found in the nebular HII, but not in the cold dense gas. 
Presumably this steepening is due to averaging out more random motions to capture the bulk flow more cleanly for the nebular HII.  
However the fact that the cold dense gas `mixing' timescale does not dramatically change suggests that the dense gas kinematics were well captured at 250~pc scales, and that we are beginning to average out actual bulk flows on 750~pc scales. 

Comparing the radial length scales over which the metallicity changes to the azimuthal scatter, \citet{Bellardini2021} found that at $R_{\rm gal} =$ 8 kpc the azimuthal scatter equaled the change in metallicity over a radial distance of $\sim$2 kpc.  And so, given a typical inward radial velocity of $\sim$10 km/s (\textit{cf.} Fig.~\ref{fig:varZcd_Vcd_250}) this radial distance would be covered in $\sim 2$ kpc/10 km/s $\approx$ 200 Myr, or roughly a dynamical time for these galaxies.  Thus, we might expect that radial flows, mixing into the local surroundings on roughly a dynamical time, are indeed able to source a significant fraction of the azimuthal metallicity scatter.

\begin{table}\caption{Best fit $\gamma$, the effective mixing speed in fractions of a dynamical time, from Eq.~\ref{eq:model} for each ISM phase at 250~pc and 750~pc pixel sizes in five of the FIRE-2 Milky Way-like galaxies used in this work}
\label{table:gamma_fits}
\centering
\begin{tabular}{rcccc}
		& \multicolumn{2}{c|}{250 pc}		& \multicolumn{2}{c}{750 pc}                \\ \hline
Name     	& $\gamma_{\rm HII}$ & $\gamma_{\rm C\& D}$ & $\gamma_{\rm HII}$ & $\gamma_{\rm C\& D}$ \\ \hline
m12b 	& 0.34 & 0.23 & 0.47 & 0.21 \\
m12c 	& 0.39 & 0.17 & 0.47 & 0.15 \\
m12f 	& 0.29 & 0.12 & 0.45 & 0.08 \\
m12i 	& 0.36 & 0.16 & 0.45 & 0.11 \\
m12m 	& 0.37 & 0.29 & 0.47 & 0.28 \\ \hline
\end{tabular}
\end{table}
\section{Impacts of the FIRE-2 Turbulent Metal Diffusion Model on Metallicity Variations}\label{sec:tmd}
\begin{figure*}
	\includegraphics[width=\textwidth]{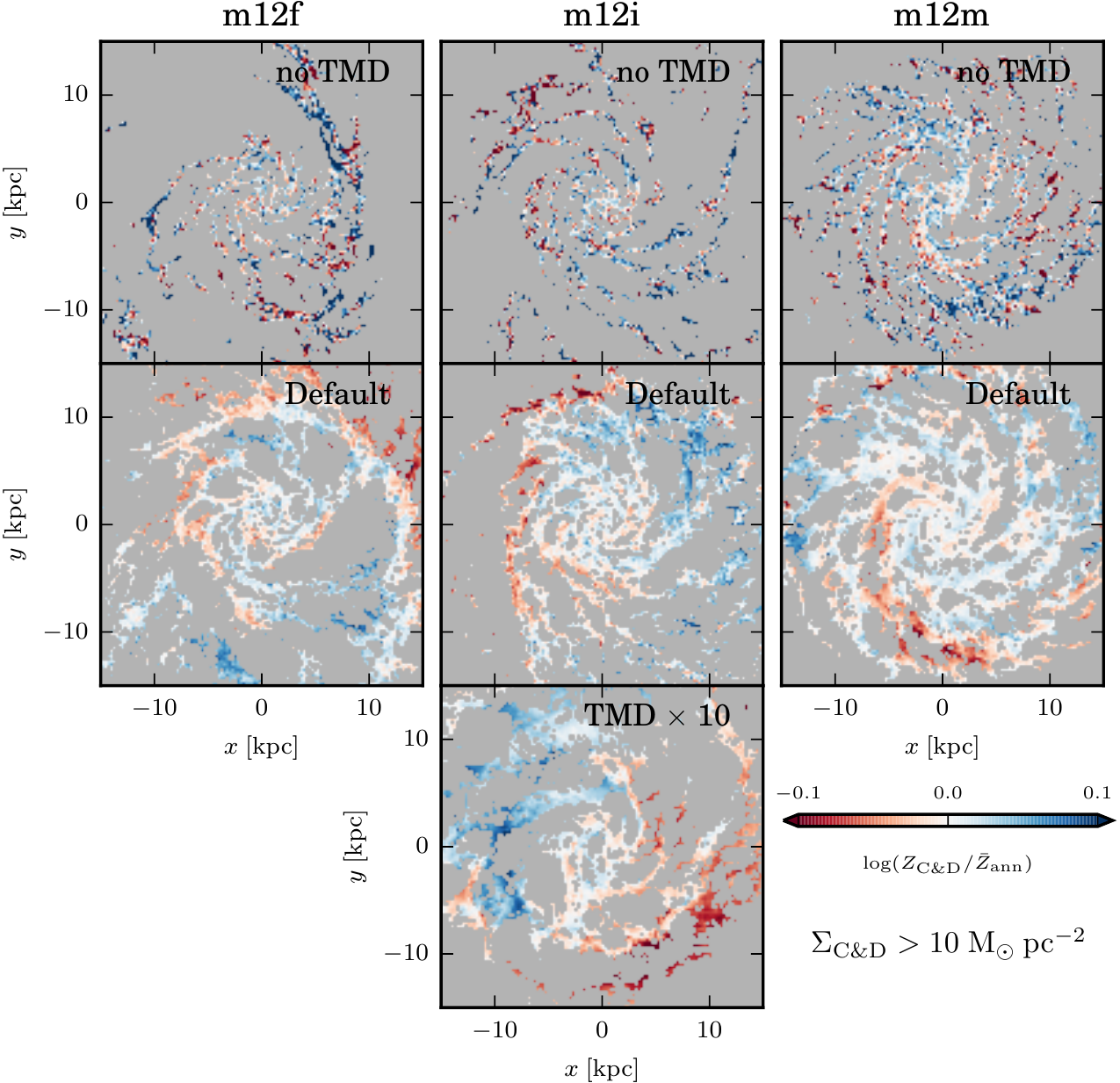}
    \caption{Comparing the effects on the cold and dense gas-phase metallicity variations, morphologically and in magnitude, of the strength of the sub-grid turbulent metal diffusion (TMD) in three FIRE-2 galaxies. Definition of cold and dense gas, and plot style as in Figure~\ref{fig:Zcd_250}.  The runs without metal diffusion result in spindly spiral arms of cold and dense gas with dramatic local variations due to local enrichment.  In one case (\textbf{m12i}) the galaxy was re-run with ten times the standard diffusion coefficient, resulting in a dramatically smoothed spatial metallicity distribution-- changing significantly only over kiloparsec scales. The spiral arms appear to act as conduits for metal transport, giving arise to azimuthal metallicity variations only in cases where there is some degree of sub-grid mixing, and when the sub-grid mixing is not so fast as to wipe out metallicity variations over kiloparsec scales.}
    \label{fig:ZcoolxCoolGas_TMD_250}
\end{figure*}
\begin{figure*}
	\includegraphics[width=\textwidth]{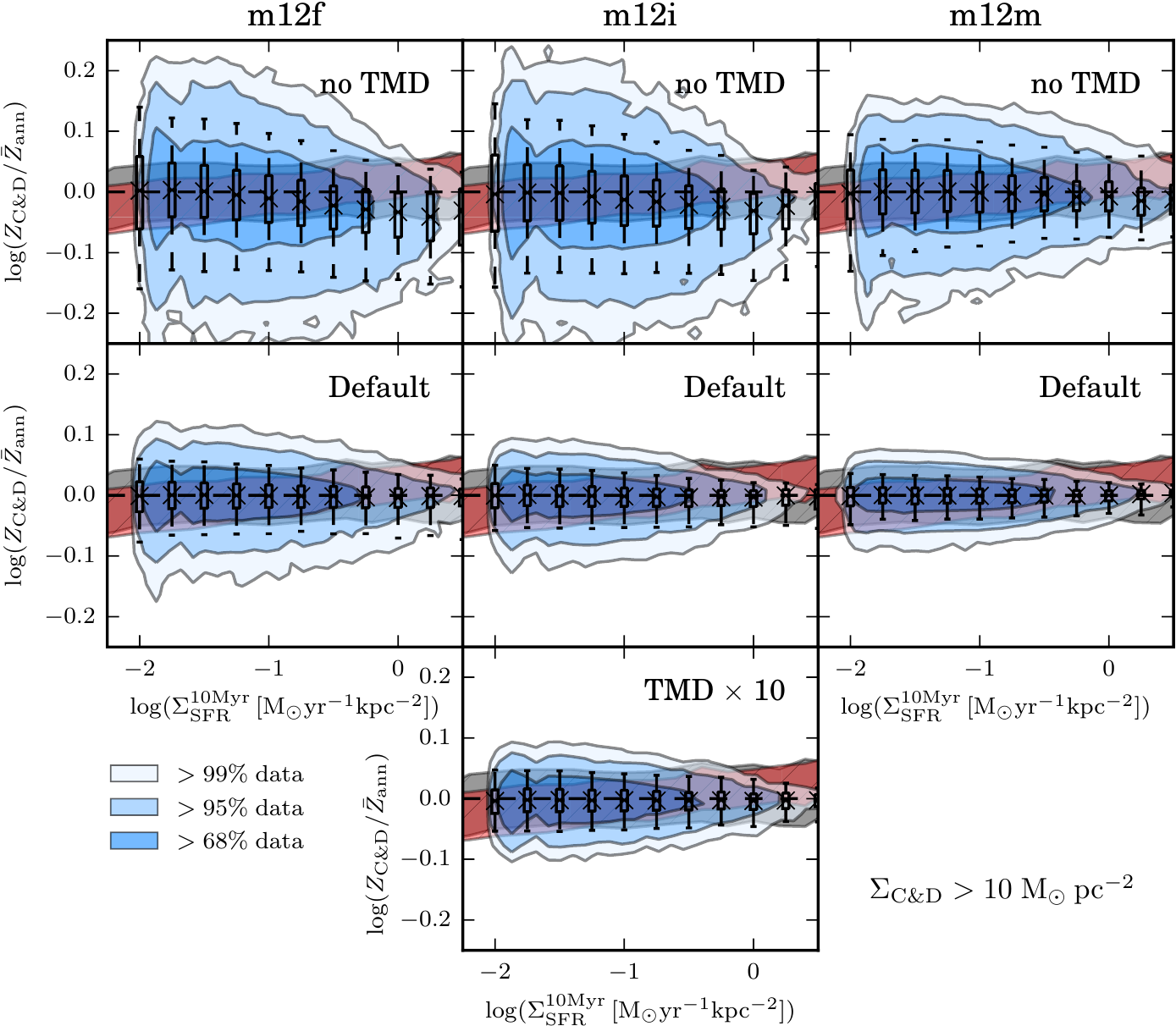}
    \caption{Comparing the turbulent metal diffusion runs (TMD) from Figure~\ref{fig:ZcoolxCoolGas_TMD_250}, in the style of Figure~\ref{fig:varZcool_SFR10_100}.  Without TMD, the azimuthal variations in gas-phase metallicity appear random, and have dramatically larger scatter at all $\Sigma_{\rm SFR}$ compared to the standard TMD runs.  In the one $10 \times$TMD \textbf{m12i} run, the scatter as a function of $\Sigma_{\rm SFR}$ does not appear to be strongly affected (perhaps slightly smaller)-- suggesting again that the strength of local azimuthal variations are not primarily driven by local enrichment or mixing, beyond the amount of both required to produce a realistic multiphase ISM.}
    \label{fig:varZcool_SFR_TMD_250}
\end{figure*}
\begin{figure*}
	\includegraphics[width=\textwidth]{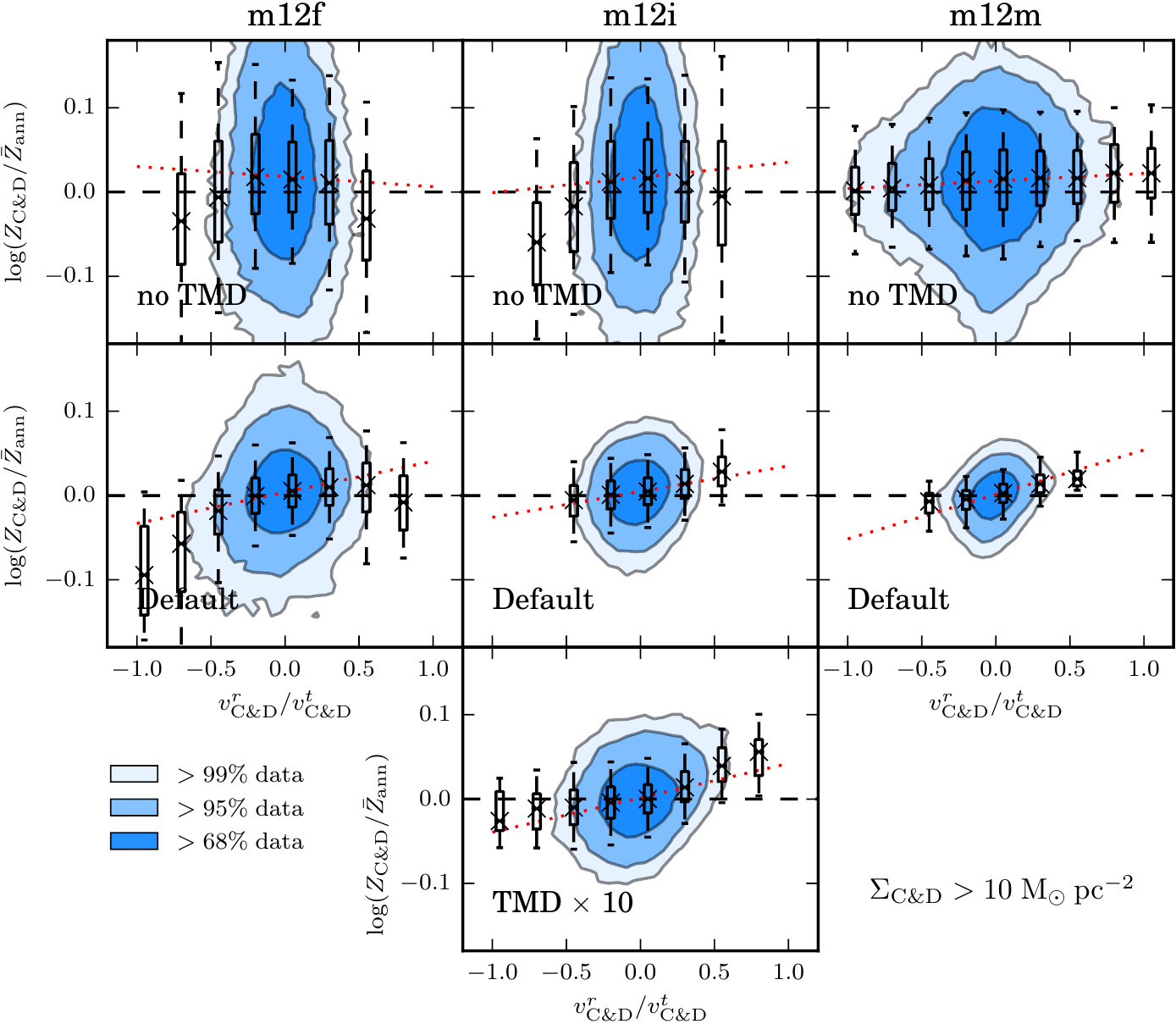}
    \caption{Comparing the turbulent metal diffusion runs (TMD) from Figure~\ref{fig:ZcoolxCoolGas_TMD_250}, in the style of Figure~\ref{fig:varZcd_Vcd_250}.  Without TMD, the azimuthal variations in gas-phase metallicity appear random, and have dramatically larger scatter at all $v^r/v^t$ compared to the standard TMD runs.  Moreover, without metal diffusion, there is not a strong, or consistent dependence of the azimuthal relative metallicity on local radial velocity. In the one $10 \times$TMD \textbf{m12i} run, the azimuthal relative metallicity as a function of $v^r/v^t$ does not appear to be significantly different from the ``standard" turbulent metal diffusion model runs (its fitted slope $\eta \approx 0.041$ being comparable with values in Table~\ref{table:eta_fits}).}
    \label{fig:varZcool_vrvt_TMD_250}
\end{figure*}

The inclusion of the sub-grid turbulent metal diffusion model manifestly alters the spatial distribution of gas phase metallicity and gas morphology on galactic scales. 
We explore the effects of the standard FIRE-2 turbulent metal diffusion model on the azimuthal metallicity variations in three galaxies (\textbf{m12f}, \textbf{m12i}, and \textbf{m12m}) that were run with and without the metal diffusion model, and one case (\textbf{m12i}) where it was also re-run with 10$\times$ the standard diffusion coefficient ($C_0 = 0.08$ vs. $0.008$, see Eq.~\ref{eq:diffusion} and \S~\ref{sec:methods}).  Tests in \citet{Bellardini2021} showed that the radial and vertical gas-phase metallicity gradients in the FIRE-2 galaxies were not systematically affected by changes to the turbulent diffusion coefficient.  Figure~\ref{fig:ZcoolxCoolGas_TMD_250} shows visually the difference between runs with and without turbulent metal diffusion: not only are the cold gas components of the spiral arms narrower and more spindly spatially without turbulent metal diffusion, but also have much larger azimuthal metallicity `variations'.  Arguably, these are not `variations' so much as an effectively random spatial distribution of which gas particles happened to have been enriched most recently by stellar feedback.  In fact, large-scale patterns in azimuthal metallicity variations following the spiral arms are not evident in runs without sub-grid turbulent diffusion.   In the case where the metal diffusion is 10$\times$ the standard value, we see that the gas morphology is slightly smoother than the default run, though within what might be expected from stochastic evolution.  
The metallicity variations, however, are smoothed out to a greater extent on the scale of a kiloparsec.  
With a order of magnitude larger diffusion coefficient, the sub-grid model effectively is operating to smooth out differences at a much larger spatial scale than across individual gas cells.

We also explore the potential connection between local star formation/enrichment and the azimuthal metallicity variations in the context of the turbulent metal diffusion model in Figure~\ref{fig:varZcool_SFR_TMD_250}. 
The overall scatter in metallicity relative to the mean in the 500~pc-wide annuli is about a factor of two larger in the non-turbulent metal diffusion runs compared to the default model, roughly in-line with the difference in scatter seen in the FIRE dwarf galaxies analyzed by \citet{Escala2018}. 
We find that without turbulent metal diffusion, unlike in the `standard' runs, the simulations do exhibit a dependence in the azimuthal variations on local star formation rate: higher local star formation rates are correlated with cold dense gas being relatively metal poor.  
A potential reason for this bias may be that metal-rich gas cells are those that have been recently affected by stellar feedback, and so metal-poor gas cells are more likely to be in a collapsing/dynamically cold state, resulting in star formation.  
Thus, averaging over 250 pc scales informs us that metal-rich pixels have been affected more by recent, disruptive feedback than metal-poor pixels, suppressing the local star formation rate.  
The one $10\times$ diffusion \textbf{m12i} run, like the standard diffusion version, does not show any dependence of azimuthal metallicity variations on local star formation rate.  
Moreover, the overall scatter in azimuthal metallicity relative to the mean is not meaningfully different than the `default' run, suggesting that the remaining scatter in metallicity is indeed driven by disk dynamics and emergent physics above the sub-grid scale.

Figure~\ref{fig:varZcool_vrvt_TMD_250} shows the dependence of the azimuthal metallicity variations on radial gas velocity, as explored in \S~\ref{sec:gasflow}.  
The runs without turbulent metal diffusion again show significantly more scatter, to the point of randomness, compared to either the default or $10\times$ cases.  
The runs lacking metal diffusion have best-fit lines that are offset from the origin, and in at least the case of \textbf{m12f} a negative slope, together indicating that the gas-flows model struggles to explain \emph{any} of the azimuthal scatter in these simulations.   
Whereas the $10\times$ metal diffusion case, the relative slope and scatter of the metallicity variations with radial velocity is in fairly close agreement with its `default' counterpart, showing a relation in-line with the gas-flows model.  
This suggests that the sub-grid mixing model does not significantly contribute to the azimuthal metallicity variations in the gas at least beyond the necessity of its inclusion to produce a realistic ISM in the first place.

Additionally, the fact that our results relating to the gas-flows model are robust to a factor of ten increase in the turbulent metal diffusion coefficient, and appear to be converged to a reasonable degree, allay some of the concerns from recent work by \citet{Rennehan2021} that the FIRE-2 sub-grid turbulent diffusion model was mixing by a factor of $\sim 20$ too low.  
Again, as the sub-grid model does not appear to affect the overall degree of azimuthal scatter in metallicity (ignoring the case of $C_0 =0$, no diffusion), the scatter must then be tied to either dynamics relating to the gas disk, or the direct impact of stellar feedback/enrichment.

\section{Discussion}\label{sec:discussion}
\subsection{Flocculent vs. Grand Design Spirals}
The FIRE-2 simulation suite predominantly includes flocculent spiral galaxies, when considering the $M_{\rm halo} \approx 10^{12}$ \msolt~MW-mass analogues near $z \approx 0$.  
However, a number of galaxies in the suite develop bars though weak and transitory, at late times \citep[][Ansar et al. \emph{in prep.}]{Debattista2019}.  
Often, the source of spiral arm structure in flocculent disks is attributed to a simple ``stochastic self-propagative star formation" model \citep{Gerola1978}, wherein the disks are found to be in some dynamical equilibrium state and there is no strong, persistent arm structure \citep{Dobbs2014}.  
On the other extreme of `arm-ness', tidal forcing from companion galaxies (or strong bars) is a strong driver of Grand Design morphology \citep{Kendal2015}.  
Observationally, \citet{Ho2017} explored the azimuthal metallicity gradients in the Grand Design spiral NGC 1365, finding a $\sim$0.2~dex jump in metallicity across the arms.  
They primarily interpreted this through the lens of a self-enrichment model, where significant star formation in the arm (at a particular radius) is able to enrich gas as it crosses the arm, highly enriching the nebular gas before it is able to mix with the lower metallicity gas in the inter-arm regions.  
Though they note the possibility of the role of radial gas flows, they estimated that radial flow velocities on the order of $\sim$20 km/s would be required to explain the metallicity jump, which the authors discounted as too large. Indeed, given the large contrast in local star formation rate surface density in Grand Design systems between arm and inter-arm regions, as compared to flocculent galaxies where such arm--inter-arm distinctions are difficult to make (like those disks analyzed here), their work clearly highlights a different regime under which azimuthal metallicity variations on the order of $\sim$0.1~dex can arise in the nebular gas.  Following work by \citet{Sanchez-Menguiano2020} explicitly comparing the magnitude of azimuthal metallicity variations between Grand Design and flocculent disk galaxies supported this notion, finding that the mean arm-interarm abundance variations were $\sim$0.015~dex higher in Grand Design versus flocculent systems.

\section{Summary \& Conclusions}\label{sec:conclusions}
In this paper, we investigated the properties of the azimuthal variations in gas-phase metallicity in six Milky Way mass cosmological zoom-in simulations from the FIRE-2 suite.  
To do so, we mapped the simulated spiral galaxies face-on (relative to their stellar disks) in a Cartesian grid with pixel sizes of 250 to 750 pc, calculating the average metallicity of those pixels in gas roughly corresponding to cold \& dense (molecular) gas, with $T < 500$~K and $\Sigma_{\rm C\&D} > 10$ \msolt~\pcsqt, and nebular HII regions, where ionized hydrogen at $T \approx 10^4$~K is in the immediate vicinity of young massive stars ($\Sigma_{\rm SFR}^{\rm 10 \, Myr} > 0$).  We also mapped local star formation rate surface densities, gas surface densities, and dynamical times to investigate the dependence of the azimuthal metallicity variations on other local quantities.

Our key findings relating to the azimuthal gas-phase metallicity variations are as follows:
\begin{itemize}
    \item Azimuthal variations in gas-phase metallicity, relative to the mass-weighted mean at a given radius, have coherent structure that follow spiral arms in galactic disks (see, \S~\ref{sec:AzZstructure} and Figs.~\ref{fig:Zcd_250} \& \ref{fig:ZHII_250}).  Specifically the variations are \emph{between} arm structures, and not between arm and inter-arm regions (Fig.~\ref{fig:arm-interarm}). The variations are not random scatter nor are they directly related to the scale of turbulent eddies.
    
    
    \item The azimuthal metallicity variations observed in these simulated spiral galaxies do not appear to depend on local star formation rates (see, \S~\ref{sec:noSFRcorrelation} and Fig.~\ref{fig:varZcool_SFR10_100}), unlike other models of enrichment involving star formation occurring during spiral arm passages.
    
    \item A significant source of the azimuthal metallicity variations appears to be radial mass transport of gas along spiral arms, alternating between inward-moving, relatively metal-poor arms, and outward-moving, relatively metal-rich arms (see \S~\ref{sec:gasflow} and Figs.~\ref{fig:Vcd_250}, \ref{fig:varZHII_VHII_250}, \& \ref{fig:varZcd_Vcd_250}).  This may be the predominant source of gas-phase azimuthal metallicity variations (or those traced by young star clusters) in flocculent late-type galaxies.
    
    \item The metallicity variations are not dependent on the strength of the particular sub-grid metal-mixing model employed by the FIRE-2 simulations (see Figs.~\ref{fig:ZcoolxCoolGas_TMD_250}-\ref{fig:varZcool_vrvt_TMD_250}).  The azimuthal scatter in metallicity appears to be driven by disk dynamics and emergent physics above the sub-grid scale.
    
\end{itemize}

Though at or very near the limit of our present ability to observationally discern relative metallicity differences ($\sim $ $\pm 0.1$~dex), azimuthal metallicity variations may present a novel approach to understanding how gas transport occurs along spiral arms.  Of special interest will be to explore the length scales over which these metallicity variations are coherent departures from their azimuthal averages, which will aid in constraining the balance between turbulent mixing and bulk transport in disks.  Moreover, azimuthal metallicity variations could be used as an independent measure to estimate in-plane disk-scale gas kinematics in low-inclination systems.  Future work exploring metallicity scatter may yet observationally confirm that spiral arms are metal freeways.

\section*{Acknowledgements}

BB is grateful for generous support by the David and Lucile Packard Foundation and Alfred P. Sloan Foundation. 
AW received support from: NSF via CAREER award AST-2045928 and grant AST-2107772; NASA ATP grant 80NSSC20K0513; HST grants AR-15809, GO-15902, GO-16273 from STScI.
The Flatiron Institute is supported by the Simons Foundation. IE acknowledges support from a Carnegie-Princeton Fellowship through the Carnegie Observatories.  This research was carried out in part at the Jet Propulsion Laboratory, which is operated by the California Institute of Technology under a contract with the National Aeronautics and Space Administration (80NM0018D0004).
We ran simulations using: XSEDE, supported by NSF grant ACI-1548562; Blue Waters, supported by the NSF; Frontera allocations AST21010 and AST20016, supported by the NSF and TACC; Pleiades, via the NASA HEC program through the NAS Division at Ames Research Center.
This research has made use of NASA's Astrophysics Data System.

\section*{Data Availability}

The data supporting the plots within this article are available on reasonable request to the corresponding author. A public version of the GIZMO code is available at \url{http://www.tapir.caltech.edu/\~phopkins/Site/GIZMO}. Additional data including simulation snapshots, initial conditions, and derived data products are available \citep{Wetzel2022} at \url{http://flathub.flatironinstitute.org/fire}.



\bibliographystyle{mnras}
\bibliography{library} 



\appendix
\section{Total Metallicity Versus Relative Oxygen and Iron Abundances}\label{sec:appendix:ZOFe}

\begin{figure*}
	\includegraphics[width=0.78\textwidth]{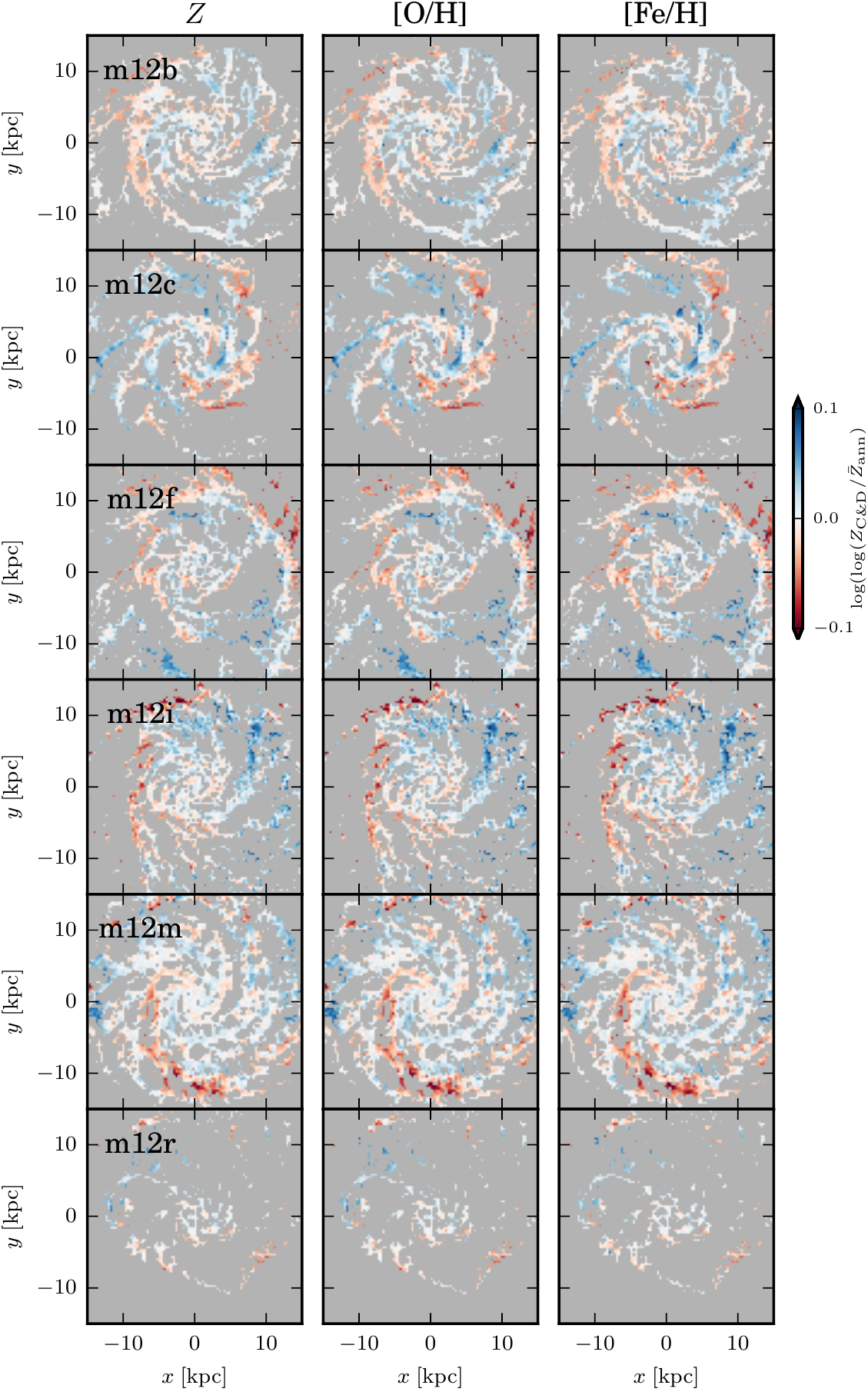}
    \caption{Spatial distribution of gas-phase metallicity, relative to median metallicity calculated within $\pm$250 pc of the galactocentric radius of each pixel, with a pixel size of 250~pc, for the cold and dense gas (where $T<500$~K, and $\Sigma_{\rm C\&D} > 10$ \msolt~\pcsqt) in the six Milky Way mass simulations analyzed here. \emph{Left to right columns}: relative total metallicity, Oxygen and Iron abundances. Very little difference in the variance, both in magnitude and spatial distribution, is seen between any of the abundances in all six galaxies.  A `barbershop pole'-like pattern is seen in all six galaxies in all three metal fields, where spiral arms appear to alternate between relatively metal rich and poor.}
    \label{fig:gals_zcd_250}
\end{figure*}

Figure~\ref{fig:gals_zcd_250} demonstrates the spatial distribution of azimuthal metallicity variations, in terms of total metallicity $Z$, as well as the local gas-phase Oxygen and Iron abundances, in all six roughly Milky Way mass simulations analyzed here.  We see immediately that there is little difference in the spatial distribution or magnitude (in dex) of the azimuthal variations between the different abundance tracers (total-$Z$, O, Fe).  That the variances are both very similar in magnitude and spatial distribution to each other despite their ostensibly different production channels (O predominantly from core-collapse SNe, and Fe from a balance of Type-Ia and core-collapse SNe) suggesting, again, that these patterns arise for reasons other than the spatial or temporal distribution of star formation (and subsequent enrichment).  And so, observationally, we would expect that the qualitative results found here would not change if Fe abundances from young massive stars were used as tracers of metallicity as opposed to gas-phase O abundances from nebular regions.


\bsp	
\label{lastpage}
\end{document}